\documentclass[twocolumn]{article}
\usepackage[utf8]{inputenc}
\usepackage{booktabs}
\usepackage{dblfloatfix}
\usepackage[status=final,nomargin,inline,lang=spanish]{fixme}
\fxusetheme{colorsig}
\FXRegisterAuthor{dg}{adg}{Dario}
\FXRegisterAuthor{rp}{arq}{Raquel}
\FXRegisterAuthor{sa}{asa}{Sergio}
\FXRegisterAuthor{en}{aen}{Enric}

\newcommand{\eg}{\emph{e.g.}, }     
\newcommand{\ie}{\emph{i.e.}, }     
\newcommand\etc{\emph{etc.}}

\title{Private Sources of Mobility Data Under COVID-19}

\usepackage{authblk}
\author[1]{Raquel Pérez Arnal}
\author[2]{David Conesa}
\author[1]{Sergio Alvarez-Napagao}
\author[1]{Toyotaro Suzumura}
\author[2,3]{Martí Català}
\author[2]{Enric Alvarez}
\author[1]{Dario Garcia-Gasulla}
\affil[1]{Barcelona Supercomputing Center (BSC)}
\affil[2]{Department of Physics. Universitat Polit\`ecnica de Catalunya (UPC$\cdot$BarcelonaTech), C. Jordi Girona, 1-3, 08034 Barcelona, Spain}
\affil[3]{ Comparative Medicine and Bioimage Centre of Catalonia (CMCiB), Fundaci\'o Institut d'Investigaci\'o en Ci\`encies de la Salut Germans Trias i Pujol,  Badalona, Catalonia, Spain}
\affil[ ]{\textit {\{raquel.perez,sergio.alvarez,dario.garcia\}@bsc.es, {\{david.conesa,enric.alvarez\}}@upc.edu, mcatala@igtp.cat, suzumura@acm.org}}

\date{July 2020}

\usepackage{cite}

\usepackage{graphicx}
\usepackage{hyperref}
\hypersetup{colorlinks,allcolors=red}

\usepackage{multirow}
\usepackage{eurosym}
\usepackage{geometry}
\usepackage{subfigure}
\usepackage{imakeidx}

\begin{document}
\maketitle

\begin{abstract}
    The COVID-19 pandemic is changing the world in unprecedented and unpredictable ways. Human mobility is at the epicenter of that change, as the greatest facilitator for the spread of the virus. To study the change in mobility, to evaluate the efficiency of mobility restriction policies, and to facilitate a better response to possible future crisis, we need to properly understand all mobility data sources at our disposal. Our work is dedicated to the study of private mobility sources, gathered and released by large technological companies. This data is of special interest because, unlike most public sources, it is focused on people, not transportation means. \ie its unit of measurement is the closest thing to a person in a western society: a phone. Furthermore, the sample of society they cover is large and representative. On the other hand, this sort of data is not directly accessible for anonymity reasons. Thus, properly interpreting its patterns demands caution. Aware of that, we set forth to explore the behavior and inter-relations of private sources of mobility data in the context of Spain. This country represents a good experimental setting because of its large and fast pandemic peak, and for its implementation of a sustained, generalized lockdown. We find private mobility sources to be both correlated and complementary. Using them, we evaluate the efficiency of implemented policies, and provide a insights into what new normal means in Spain.
 
\end{abstract}

\section{Introduction}

COVID-19 has produced an unprecedented change in mobility \cite{Bonaccorsi-2020-Economicandsocial,Gatto-2020-Spreadanddynamics}. Governments have implemented restrictive measures to contain the pandemic, focused on reducing social contacts~\cite{Beria-2020-Presenceandmobilit}. This is a direct effort towards controlling the COVID-19 spread, at a stage where the test and trace strategy is impractical~\cite{Kraemer-2020-Theeffectofhuman}. Mobility restriction measures range from the closure of schools and large gatherings, to the complete lockdown of the population and economic hibernation. Different countries have implemented different measures, varying their severity and duration \cite{Hegedus-2020-ECMLCoviddashboard}. Lifting these restrictions will lead societies to a new normality, which is still unclear how much will resemble the old normality. At this point it seems likely that societies will transition to a new basal mobility structure \cite{Funk-2009-Thespreadofawaren}. One we need to analyze and understand the sooner the better.


There are many public sources of mobility data which can be used to measure the physical distancing of population during the COVID-19 crisis \cite{MovindEstado,FlightData,Oliver-2020-Mobilephonedatafo}. Unfortunately, most of these sources provide limited insights, focusing on a movement modality (\textit{e.g.,} public transport occupancy, public bike system usage, road densities). An alternative to public data is the data provided by private technological entities, which have access to a high volume of information through applications installed in mobile phones \cite{Spyratos-2019-Quantifyinginternat}. Even though the segment of population that has these applications installed is not a perfect reflection of society (\textit{e.g.,} young segments of society are over-represented while elders and kids are under-represented) it is large enough to provide reliable estimations. At the moment, data provided by private entities represents the most reliable public source of information to explore the big picture from the perspective of people~\cite{Buckee-2020-Aggregatedmobility}. Such sources of data include the ones provided by Google, Facebook and Apple, as described in Section \ref{sec:data}.

The main issue when working with private data sources is the imperfect knowledge regarding its nature. For privacy preserving reasons, raw data is never provided. Instead, data goes through heavy pre-processing and anonymization procedures \cite{Herdagdelen-2020-Protectingprivacyi}. Additionally, the conditions under which data is collected are not fully transparent either, as baselines and contextual information are often missing. To partially overcome these issues, in this work we investigate the relation between the different private data sources, and how can they be used complementary to provide a better understanding of mobility.

For our analysis, we focus on the case of Spain. The COVID-19 pandemics in Spain, and the political measures taken to control its spread in the country, provide an appropriate experimental setup. Spain implemented a complete and sudden lock-down on March 15, 2020, while the first mild restrictions were put in place less than a week before (progressive closure of schools and banning of large gatherings) \cite{Hegedus-2020-ECMLCoviddashboard}. After the pandemic peaked in April, Spain gradually recovered mobility and services over the span of two months (May and June). We will study Spain's demographics and consider the relation between restriction policies, social behavior and pandemic evolution. This could be helpful for reacting to future mobility crisis. We will also study its progression towards a new normality in comparison with the old one. This could be helpful for the adaptation of mobility policies to the new social setting.


The rest of the paper is structured as follows. First, we describe the data used in this work in \S\ref{sec:data}. Then we review the social context of the study in \S\ref{sec:spain_dem}, including both the timeline of implemented policies, and an overview of Spain demographics. Most of our analysis takes place in \S\ref{sec:analysis}. This includes a general study of mobility trends for all regions and data sources (\S\ref{sec:general_trend}), a discussion on the anomalies observed (\S\ref{sec:peaks}), an analysis on the daily trends (\S\ref{sec:daily_trend}) and some insights on the new normality (\S\ref{sec:new_norm}). Finally, we review our conclusions in \S\ref{sec:concl}.


\section{Data Sources}\label{sec:data}

For this work we have considered the data published by the Facebook Data for Good program (FDG) \cite{unknown-unknown-FacebookDataforGo}, by the Google mobility assessments \cite{Bavadekar-2020-GoogleCOVID-19Comm}, and by Apple \cite{Apple-unknown-MobilityTrendsRepo}. At the time of our analysis Apple only provided one single mobility metric, the number of direction requests made from the Apple Maps application. In practice, this measure was too different from the other sources as to be directly compared. Also, while it was possible to assess the anonymization effect on the data for FDG and Google, we have not been able to do the same for Apple. For these reasons, we disregard Apple data in this analysis.

Private mobility indices are typically provided at a certain level of aggregated granularity. With administrative level 1 being country, level 2 in Spain corresponds to region (17 Autonomous Communities or CC.AA.), and level 3 to province (50 of them in Spain). As we will see next, the Google data is provided at level 2 (regions), while Facebook data is provided at level 3 (province), a smaller geographical granularity. This will be aggregated to level 2 on our work. 

\subsection{Facebook Data for Good}\label{sec:facebook}

 Facebook Data for Good presents multiple data sets of mobility, all of them with a general description of how they are obtained. There are two main data sets of particular relevance for the purpose of this work: a remain-in-tile index and movement-between-tiles index~\cite{Maas-2019-FacebookDisasterMa}. Both are based on GPS data from a sample of users that have activated the tracking system on their mobile phones, dividing the space in level 16 tiles (squares of roughly 500~$\times$~500 m). The first one, remain-in-tile, provides the percentage of people that remains in the same tile, computed as the total ratio of mobiles providing signal which do not change of tile during a whole day. The second one, movement between tiles, estimates mobility by computing how many different tiles are visited by the sample of people, compared with the same number during the same day of the week previous to the pandemics (February 2020) \cite{Herdagdelen-2020-Protectingprivacyi}.
 
 For the rest of this work, we will use the \emph{remain-in-tile} index, as it provides a more pure measurement. Notice the remain-in-tile is an absolute measure (\eg it does not depend on a baseline). This makes it straightforward to interpret, and a good candidate for counting people who are following confinement, as long as the population who has accepted to be pinned represents a good sample of the population.  
 The anonymization of data prevents us from evaluating the fit between the data sample and the overall population distribution. A certain amount of bias is to be expected, as the penetration of smartphone and Facebook varies significantly among cohorts. As with Google, the elder population and the younger population could be slightly underrepresented, which may entail a certain bias. Nonetheless, the size of the sample lead us to believe that the data can provide a reliable picture.
 
 We assess Facebook's sample size in two ways. First, by considering Facebook market penetration in the smartphone market, which is around 50-60\% \cite{IABspain-2019-EstudioAnualMobile} with smartphones being available for 70\% of the population.\dgnote{find and cite a source for this number} The subsample with an activated tracking system only needs to be around 10\% to have a sample of 4\% of the total population. Even 1\% of the population would be a very large sample in any kind of poll. Second, data of active Facebook users is also available from Facebook geoinsights maps and indicates that 1\% is indeed the typical order of magnitude of the sample. The fact that Facebook requires a minimum of 300 active users to provide data (for anonymity reasons) allows us to validate the size of the sample. For example, the province of Spain with the smallest population is Soria, with roughly 90,000 people. This province and has never pined less than 300 people for the whole period under study (\ie there is data for Soria for all days). This means that the sample in Soria is always above 0.5\% of the population. We have no reason to suspect a lower coverage on the other provinces of Spain.

 Facebook data is provided at level 3 (province) granularity. In order to compare with Google data (which is only available at level 2), we need to aggregate it. Any level 2 region is a sum of one or more level 3 provinces. To compute the level 2 data from level 3 values we use the average of provinces weighted by their population. By doing so we obtain the ratio of people that remain in tile at level 2 aggregation for Facebook.

\subsection{Google}\label{sec:google}

On April 2, 2020, Google released its COVID-19 Community Mobility Reports~\cite{Bavadekar-2020-GoogleCOVID-19Comm}. Since then, Google periodically releases anonymyzed mobility data~\footnote{\url{http://google.com/covid19/mobility}}, organized in a set of categories (\eg retail, recreation, groceries, pharmacies, parks, transit stations).
This data is always provided at administrative level 2 granularity (region). In addition to those, Google also releases a Residential and a Workplace measure, estimating how much time is spent at those places.
In the case of Residential, it is based on the average number of hours spent at the place of residence for each user within a geographic location.
This data is collected from those users opting in to Location History, and is processed using the same algorithms used for the detection of user location in Google Places, offering an accuracy of around 100m in urban areas \cite{Rodriguez-2018-Googletimelineaccu}. Google Mobility Reports releases relative information for every day of the week.
That is, mobility with respect to a baseline: the first five weeks of the year, from January 3 to February 6.

Given the large number of users of Google Maps there is no doubt that the sample size of Google is also large compared with any standard polling sample. However, the fact that Google does not provide an absolute value in the way Facebook does presents researchers with serious challenges. The most important is that the baseline can be indeed corrupted by local festivities or changes in the normal baseline due to large-scale celebrations. As we will see, lack of access to the baseline prevents the direct comparisons of regions, and even days, and can complicate its interpretaion. 

Google's residential index, which indicates the estimated time spent at the residence, should strongly correlate with Facebook's remain-in-tile. \ie a large fraction of people remaining at home should increase both indices. Their direct comparison should provide context for their interpretation, and insights on their applicability.

\section{Social Context}\label{sec:spain_dem}

\begin{table*}[ht]
    \centering
\begin{tabular}{lrrrr}
\toprule
               Region & Pop. density & Population &   \textgreater 50k inhabitants &  Attack rate / 10$^5$ inh \\
\midrule
            Andalucía & 96 & 8,446,561 &        50.82 &          200 \\
               Aragón & 28 & 1,324,397 &      55.19 &           533\\
             Asturias & 96 & 1,019,993 &      60.75 &          238 \\
            Canarias & 298 & 2,220,270 &     54.26 &          114\\
            Cantabria & 109 & 581,949 &     38.55 &           405 \\
      Castilla y León & 26 & 2,402,877 &      44.03&          1,110 \\
   Castilla-La Mancha & 26 & 2,038,440 &       27.40 &          1,102\\
             Catalunya & 237 & 7,609,499 &       53.90 &          762 \\
             Comunitat Valenciana & 215 & 4,998,711 &       45.50 &          300 \\
             Euskadi & 302 & 2,181,919 &       46.53 &          672 \\
          Extremadura & 26 & 1,062,797 &       28.66 &          535 \\
              Galicia & 91 & 2,698,764 &       36.71 &          402 \\
       Illes Balears &  240 & 1,198,576 &     40.60 &          199\\
        La Rioja & 62 & 314,487 &       47.71 &           1,274 \\
               Murcia & 132 & 1,494,442 &      55.85 &          168 \\
               Madrid & 833 & 6,685,471 &      85.73 &          1,086 \\
              Navarra & 63 & 652,526 &       30.82 &           1,218 \\
           
\bottomrule
\end{tabular}
\caption{Population density, population, percentage of population living in cities with more than 50,000 inhabitants and attack rate (total cases) of COVID-19 for the 17 Spanish regions on the first of July }
    \label{tab:ccaa_socioeconomic}
\end{table*}

The first cases COVID-19 in Spain were reported on January 31st, in the Canary Islands. For roughly a month, all detected cases were imported from other countries. The first cases of community transmission (\ie source of contagion unknown) were diagnosed on the February 26. In March is where the scope of our analysis begins. The number cases reached 999 on the 9th. At this point a few several regional governments took the first generalized measures, with the closure of schools. This escalated until March 14, when the Spanish government declared the state of alarm. This directly imposed movement restrictions on the whole population, allowing only essential activities and work related journeys. These restrictions where reinforced on March 29 with the implementation of a \textit{hard-lockdown}, which suspended all mobility not related to essential services. This lasted until the April 12, when work trips were approved again. The first generalized de-confinement measures were adopted two weeks later, on April 26, when kids under 14 (and a tutor) were allowed outdoors for an hour. On May 2nd, small businesses were allowed to open through previous appointment. This de-confinement process continued progressively until June 21, when the last mobility restrictions in Spain were lifted. Our analysis ends on June 27.

At administrative level 2, Spain is composed by 17 regions and two autonomous cities (Ceuta and Melilla). The latter are not included in the analysis because their distinct nature would require a separate study. Even so, the 17 regions of Spain differ significantly in socio-economic factors, as shown in Table~\ref{tab:ccaa_socioeconomic}. Two regions (Catalonia and Madrid) include the largest metropolitan areas, and had the most COVID-19 cases in absolute terms. The rest of regions can be categorized as being dense or sparse (as determined by the population density), and as having their population centered on urban areas or not (as determined by the percentage of people living in cities with 50K or more inhabitants). Cantabria and Aragon are examples of the most uncommon cases, the first being fairly dense, with population centered on small towns, and the second being sparse, with population centered on cities. Extremadura is a prototypical sparse and rural region, while Madrid is extremely dense and urban.

\section{Analysis}\label{sec:analysis}

In our analysis, we focus on Facebook's remain in tile index, and on Google's residential index. The former is scaled (multiplied by 50) to approximate Google's scale and facilitate the interpretation of figures. We have no means to assess the relation between the scale of both indices, which is why we avoid interpreting the relative volume between indices. Nonetheless, as we will see in Figure~\ref{fig:monthly_trends_1}, our scale provides a good approximation.

\subsection{General Region Trend}\label{sec:general_trend} 

\begin{figure*}[h!]
        \centering \subfigure{\includegraphics[width=.32\textwidth]{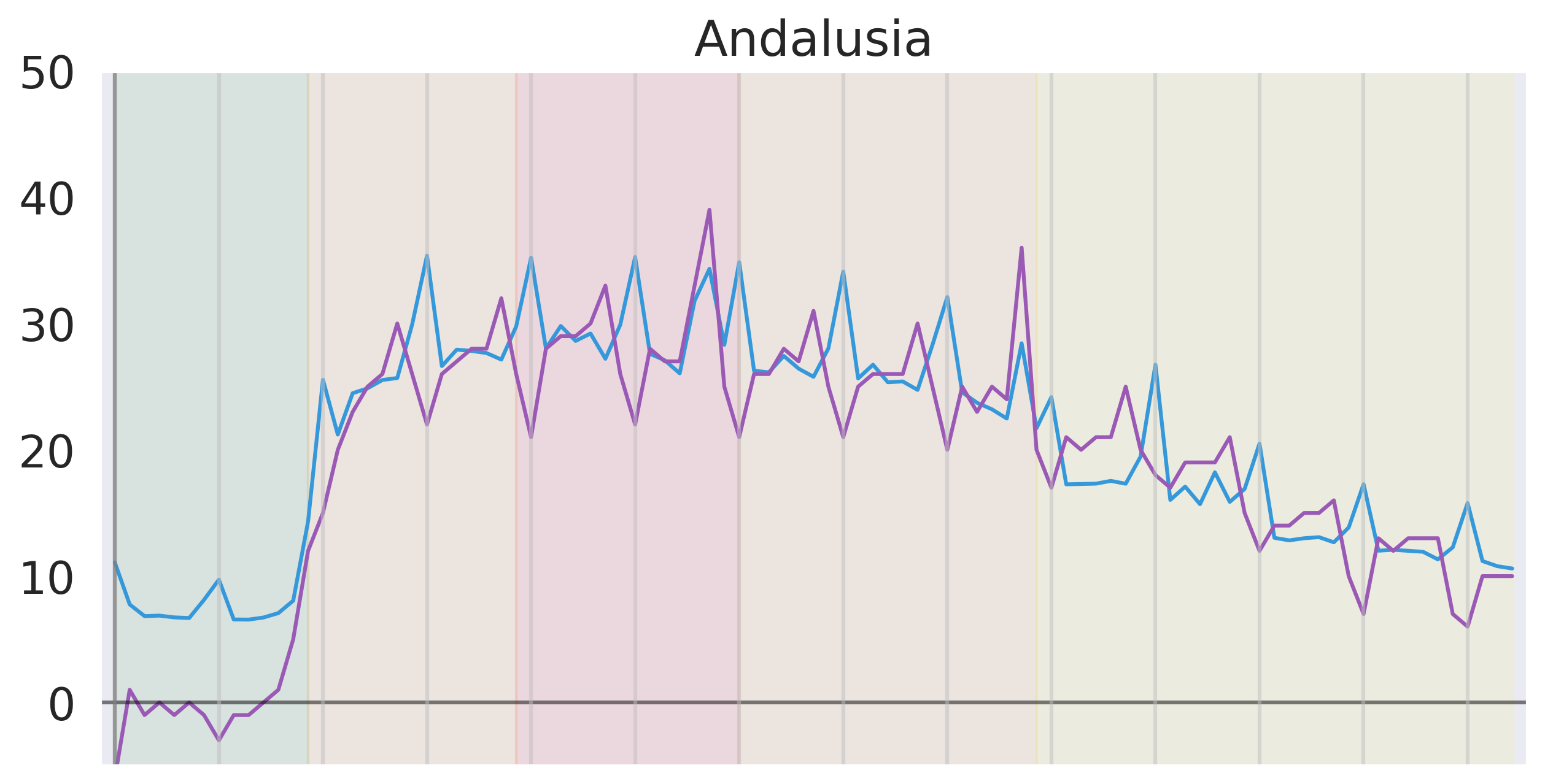}}
    \subfigure{\includegraphics[width=.32\textwidth]{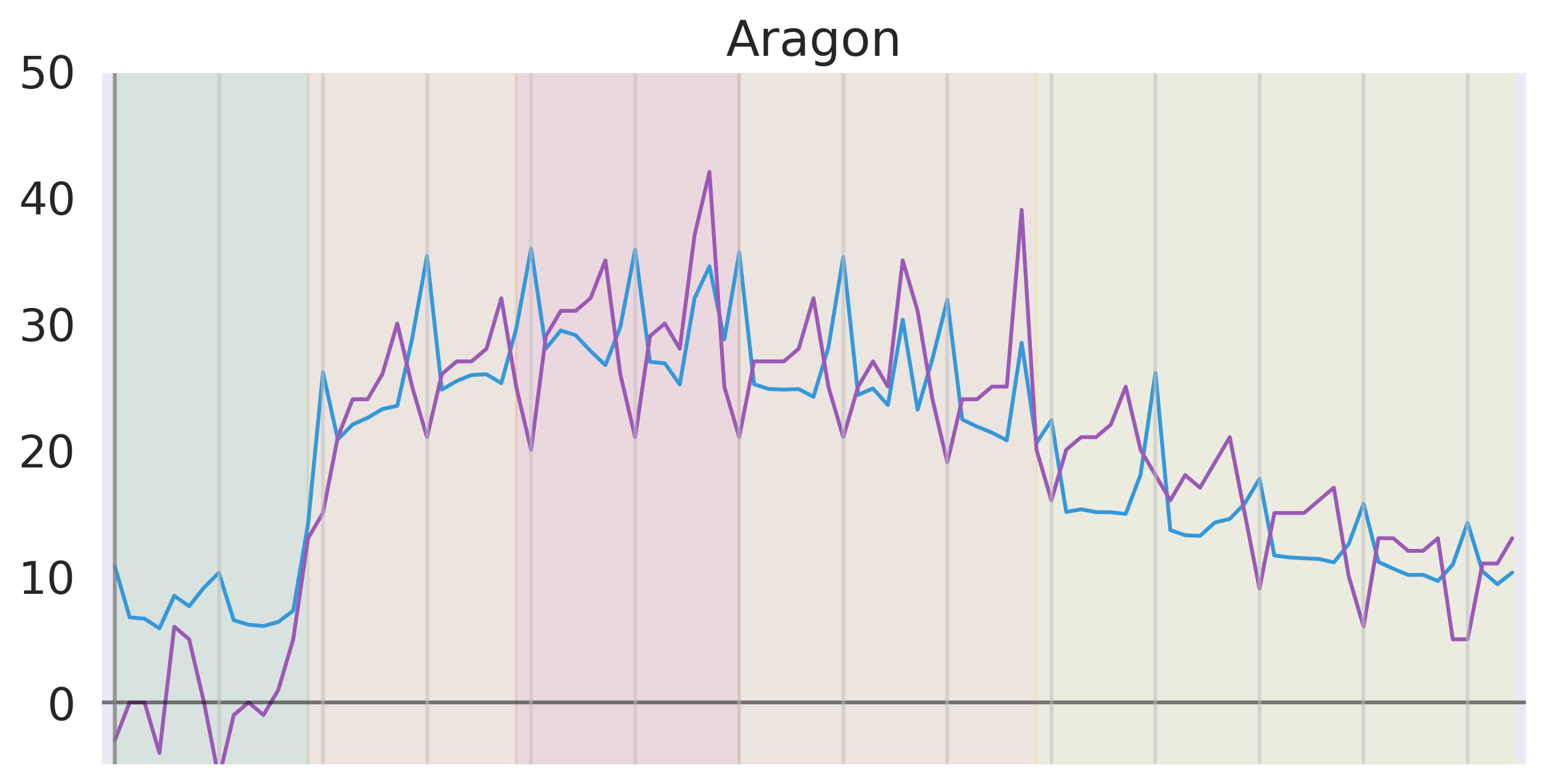}}
    \hfil
    \vfill

    \subfigure{\includegraphics[width=.32\textwidth]{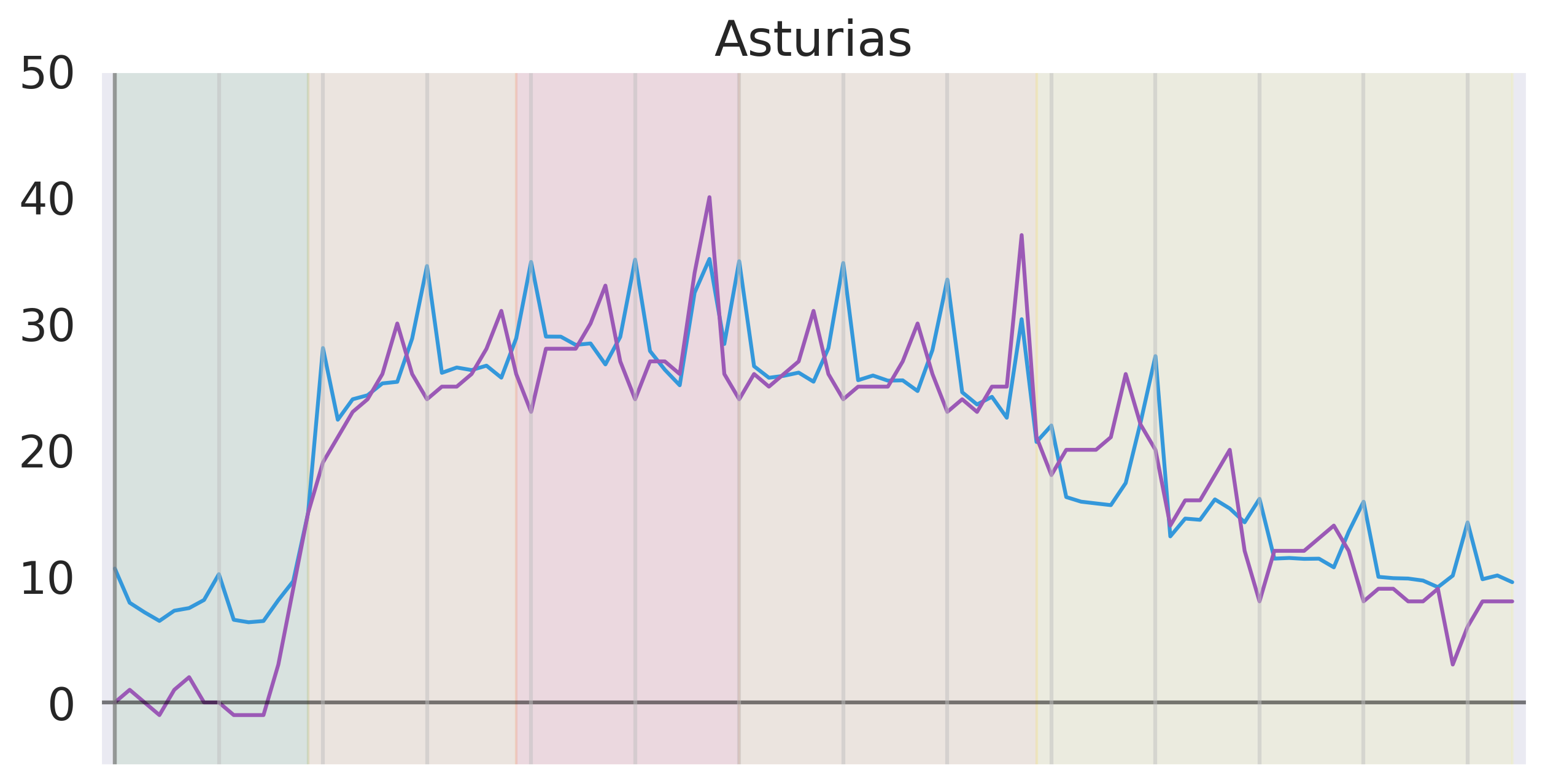}}
    \subfigure{\includegraphics[width=.32\textwidth]{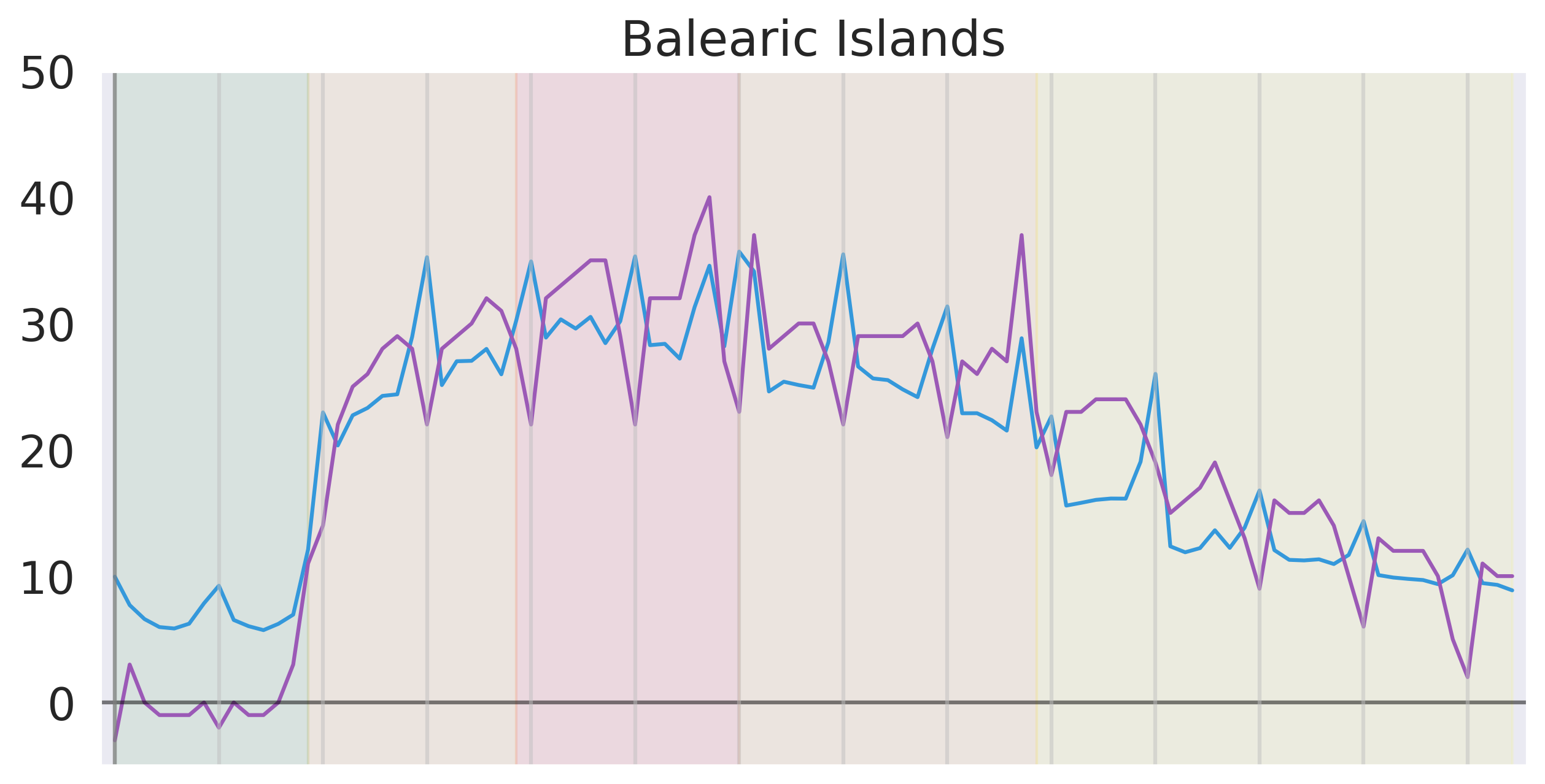}}
    \subfigure{\includegraphics[width=.32\textwidth]{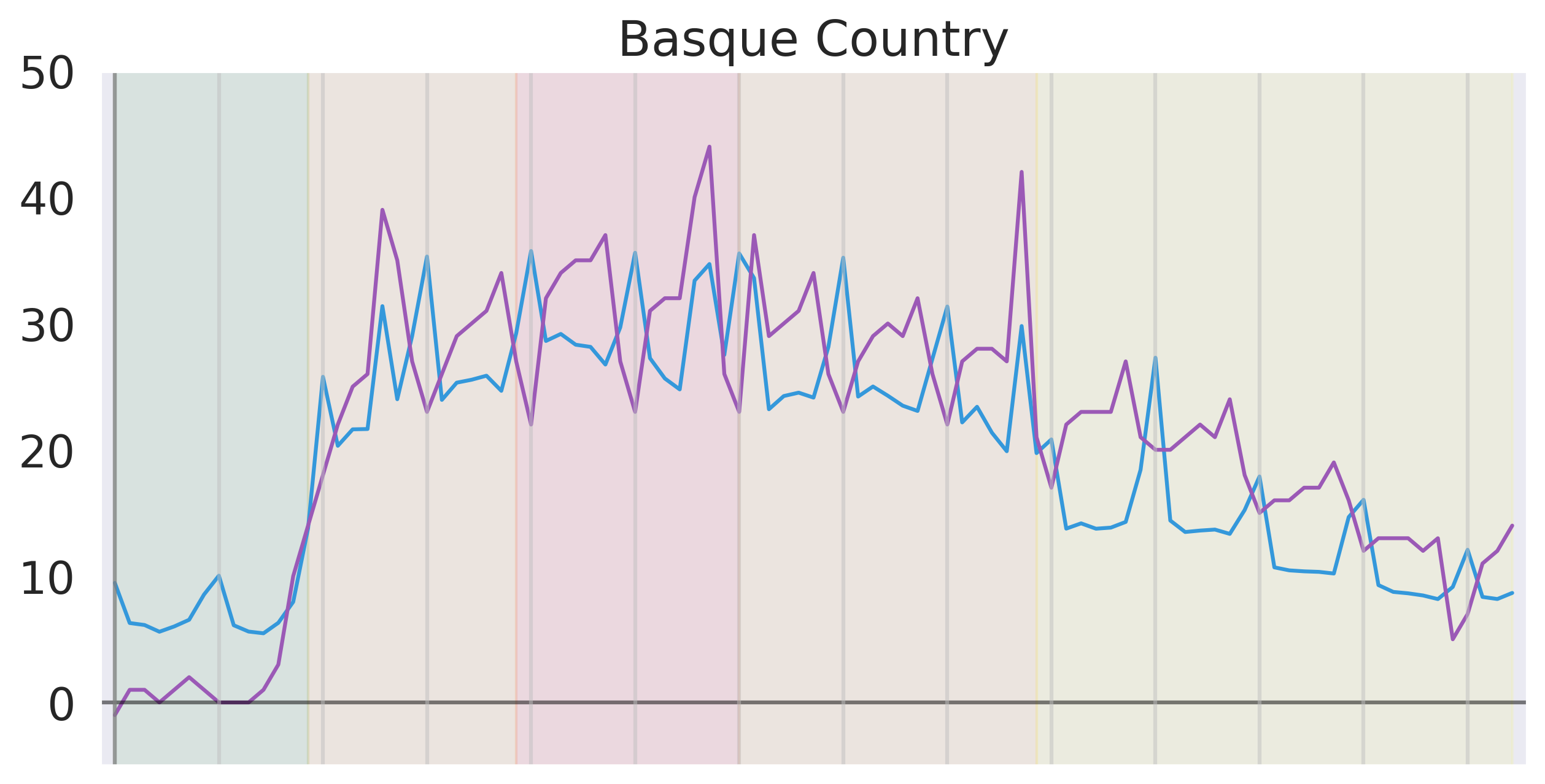}}
    \subfigure{\includegraphics[width=.32\textwidth]{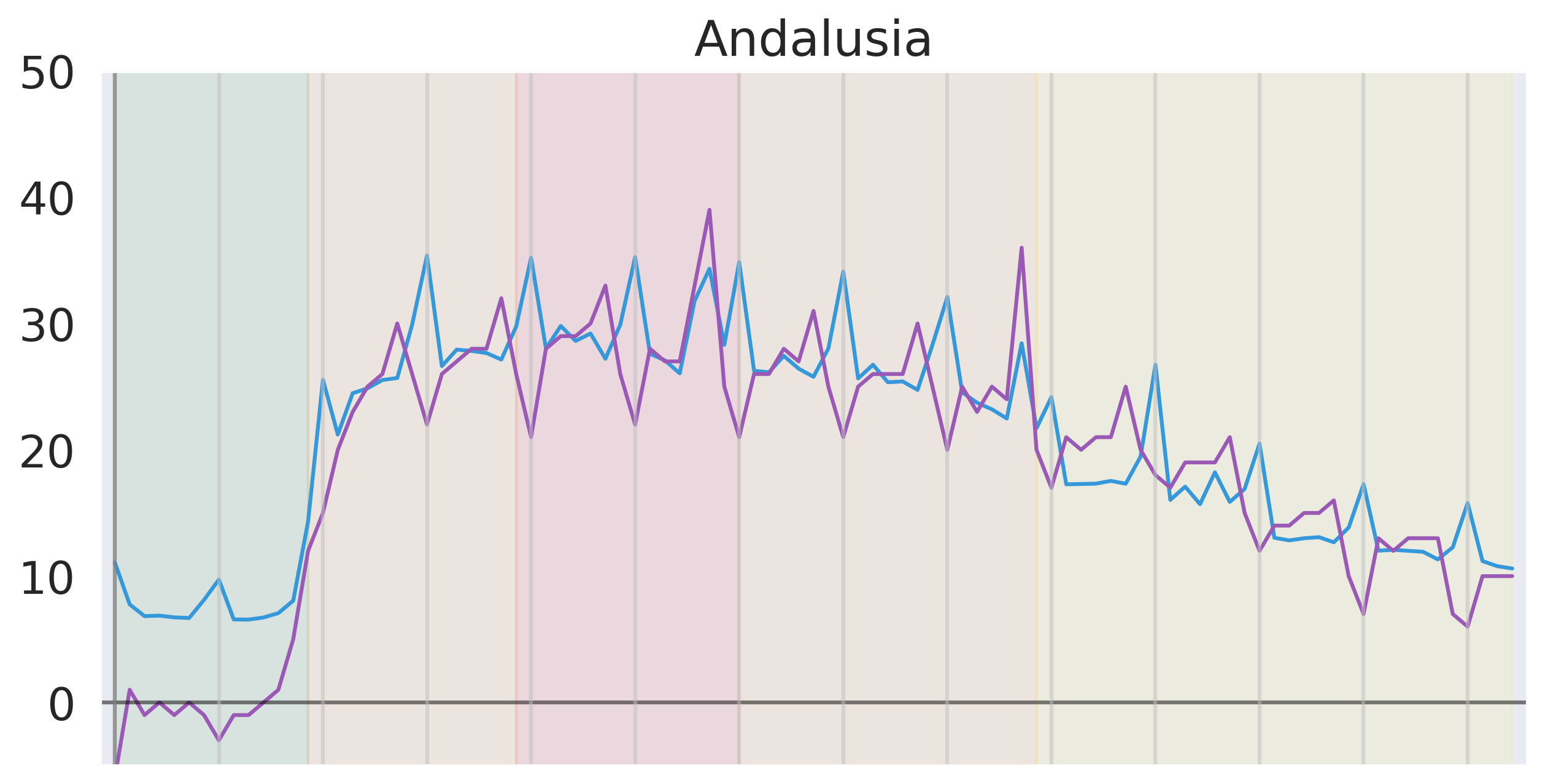}}
    \subfigure{\includegraphics[width=.32\textwidth]{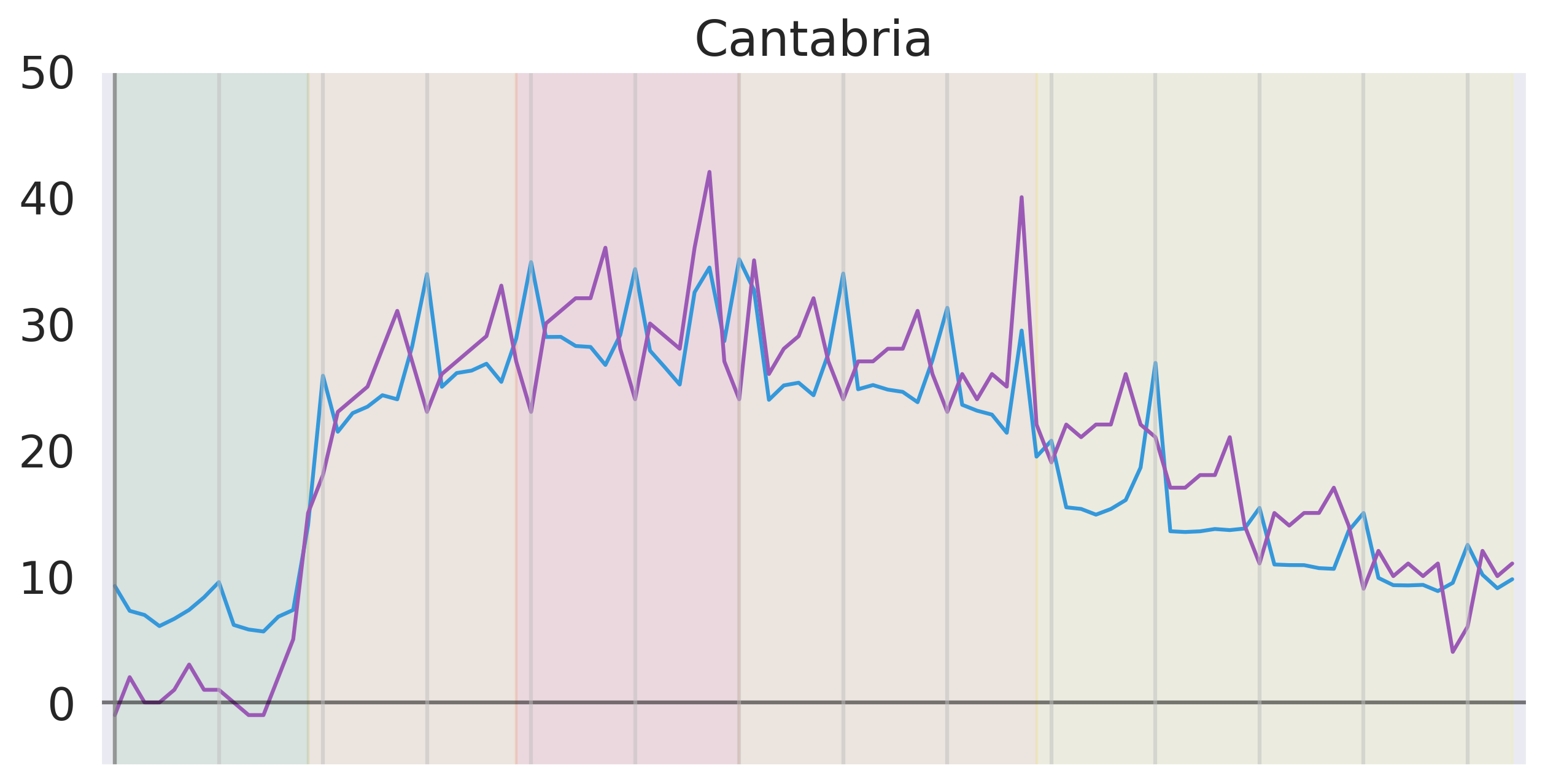}}
    \subfigure{\includegraphics[width=.32\textwidth]{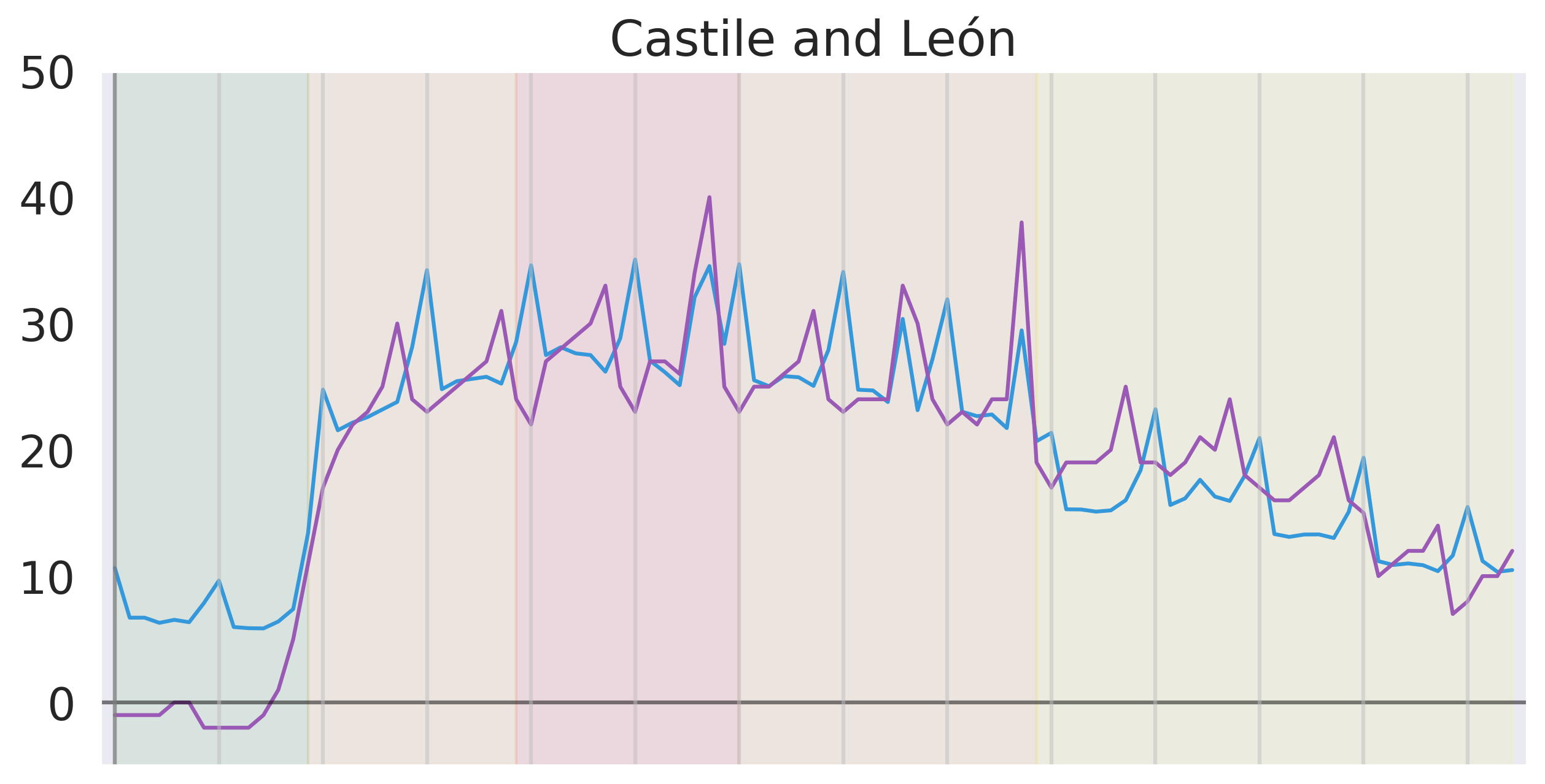}}
    \subfigure{\includegraphics[width=.32\textwidth]{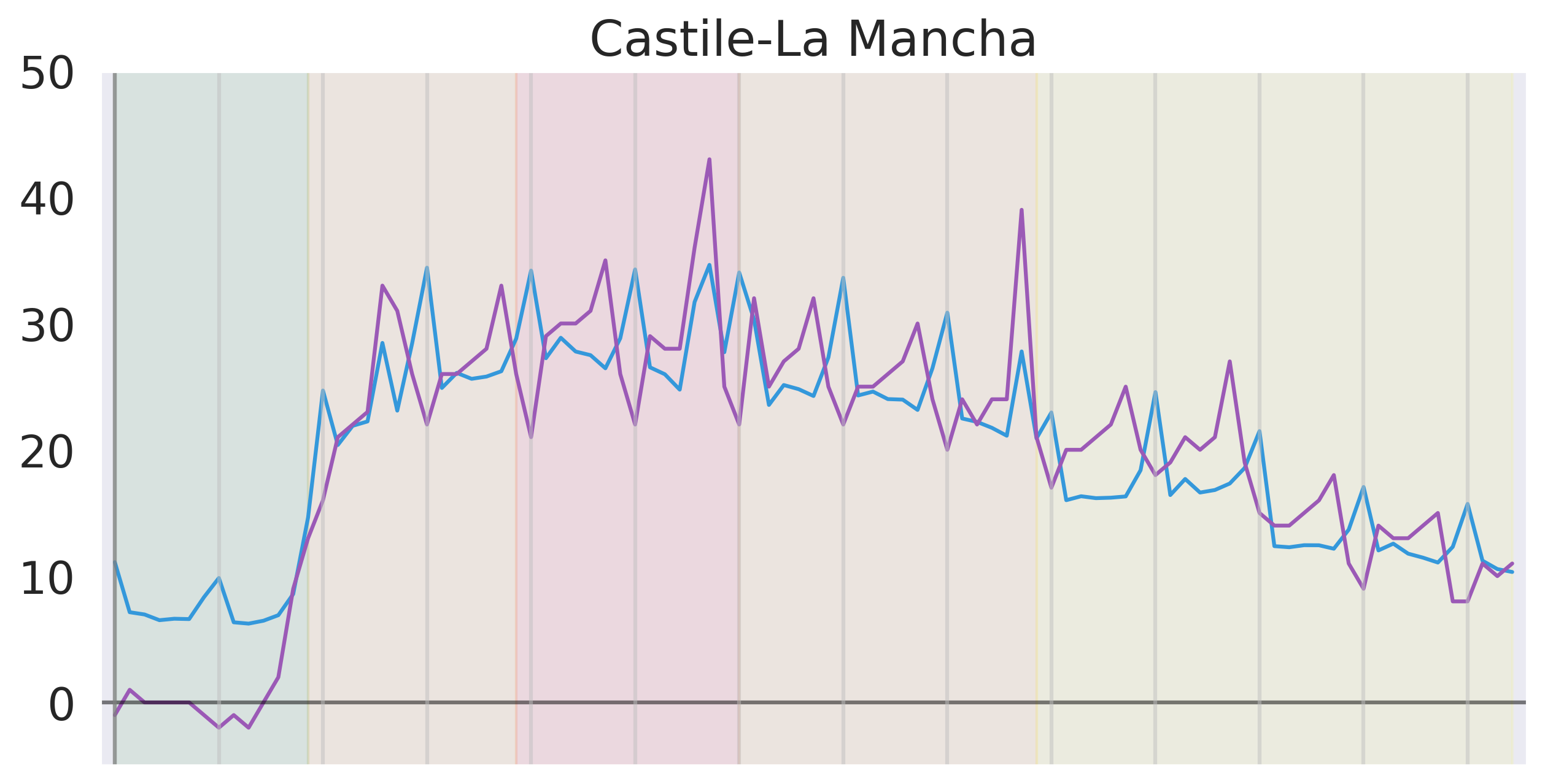}}
    \subfigure{\includegraphics[width=.32\textwidth]{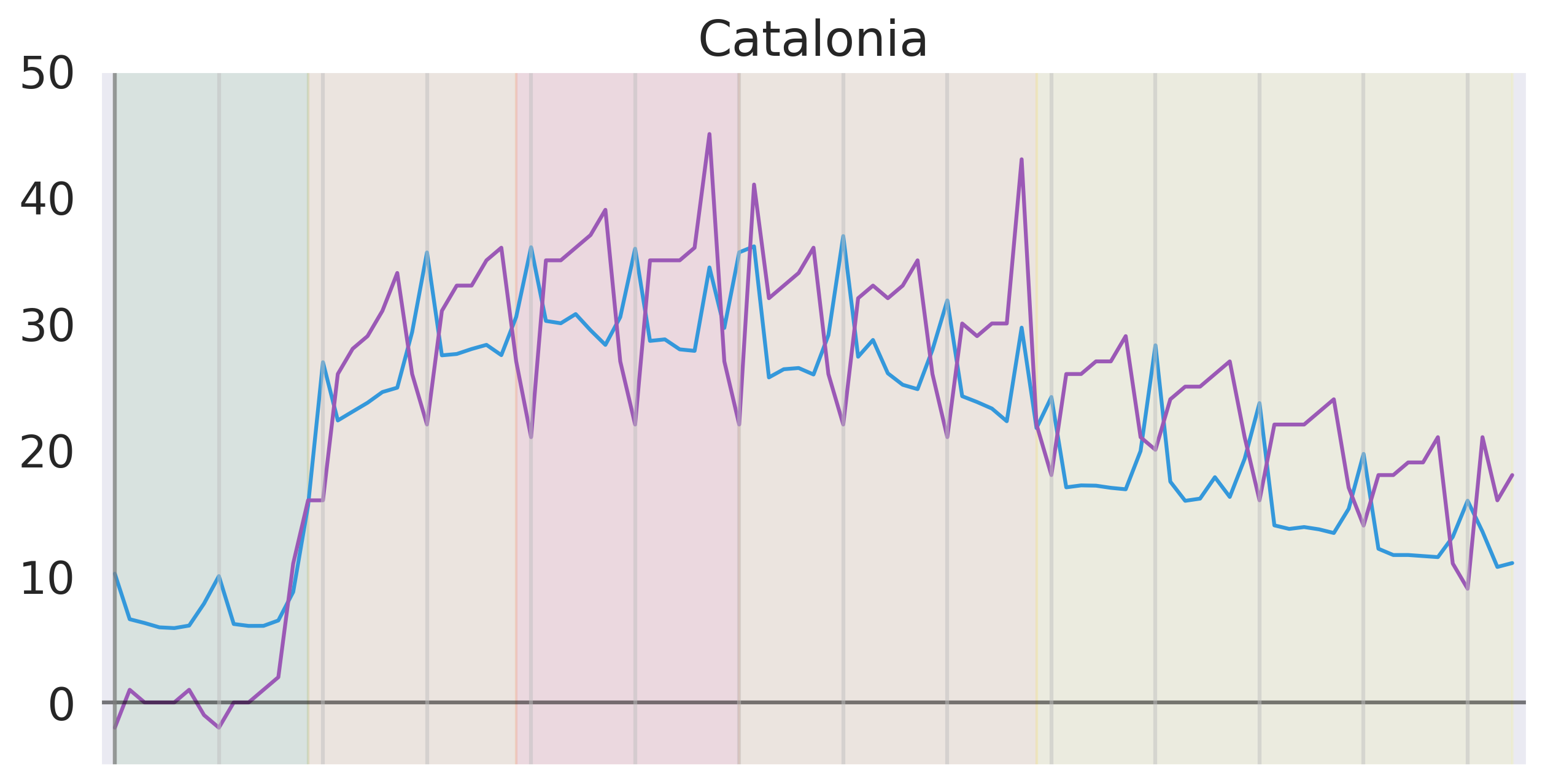}}
    \subfigure{\includegraphics[width=.32\textwidth]{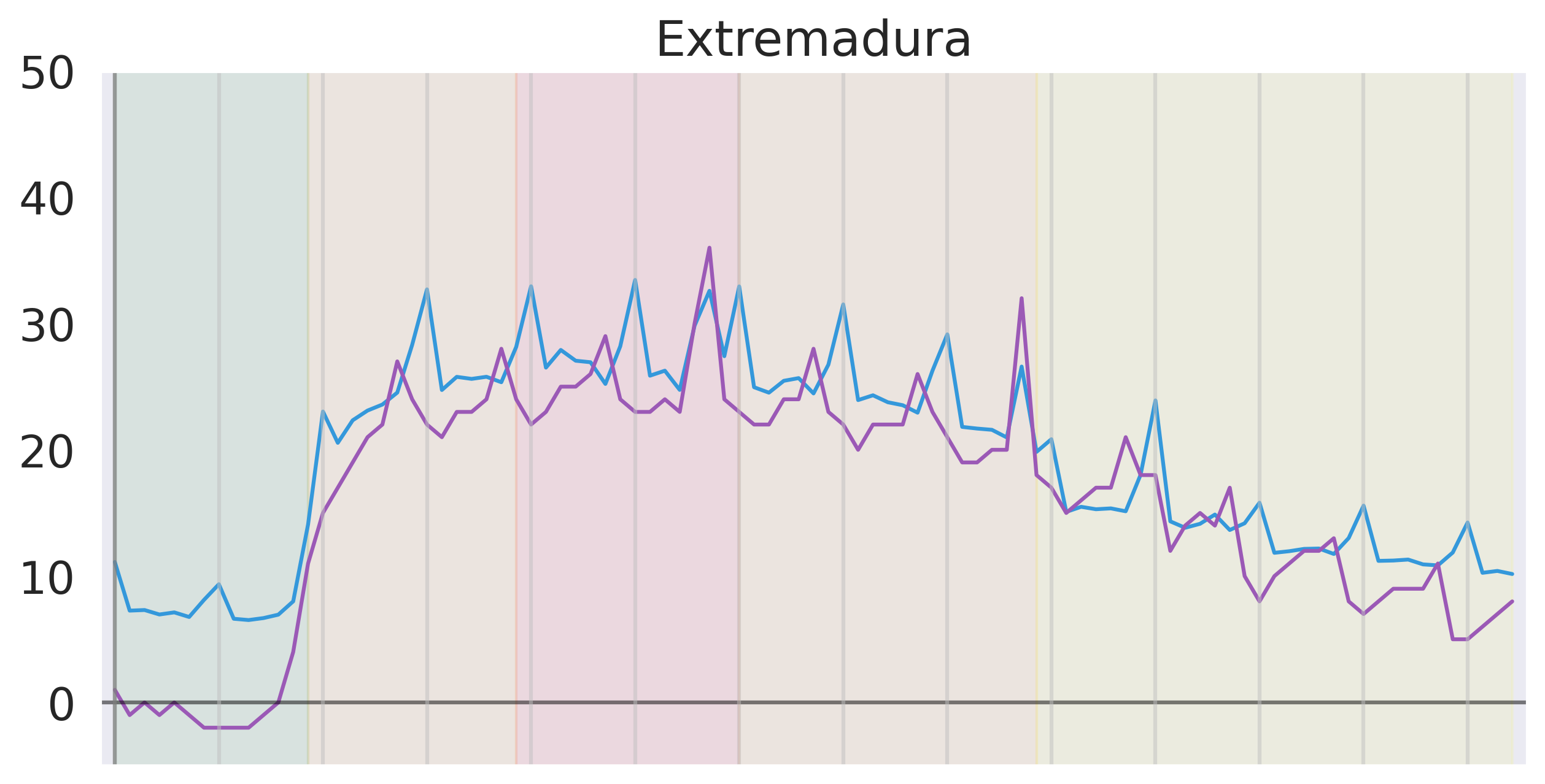}}
    \subfigure{\includegraphics[width=.32\textwidth]{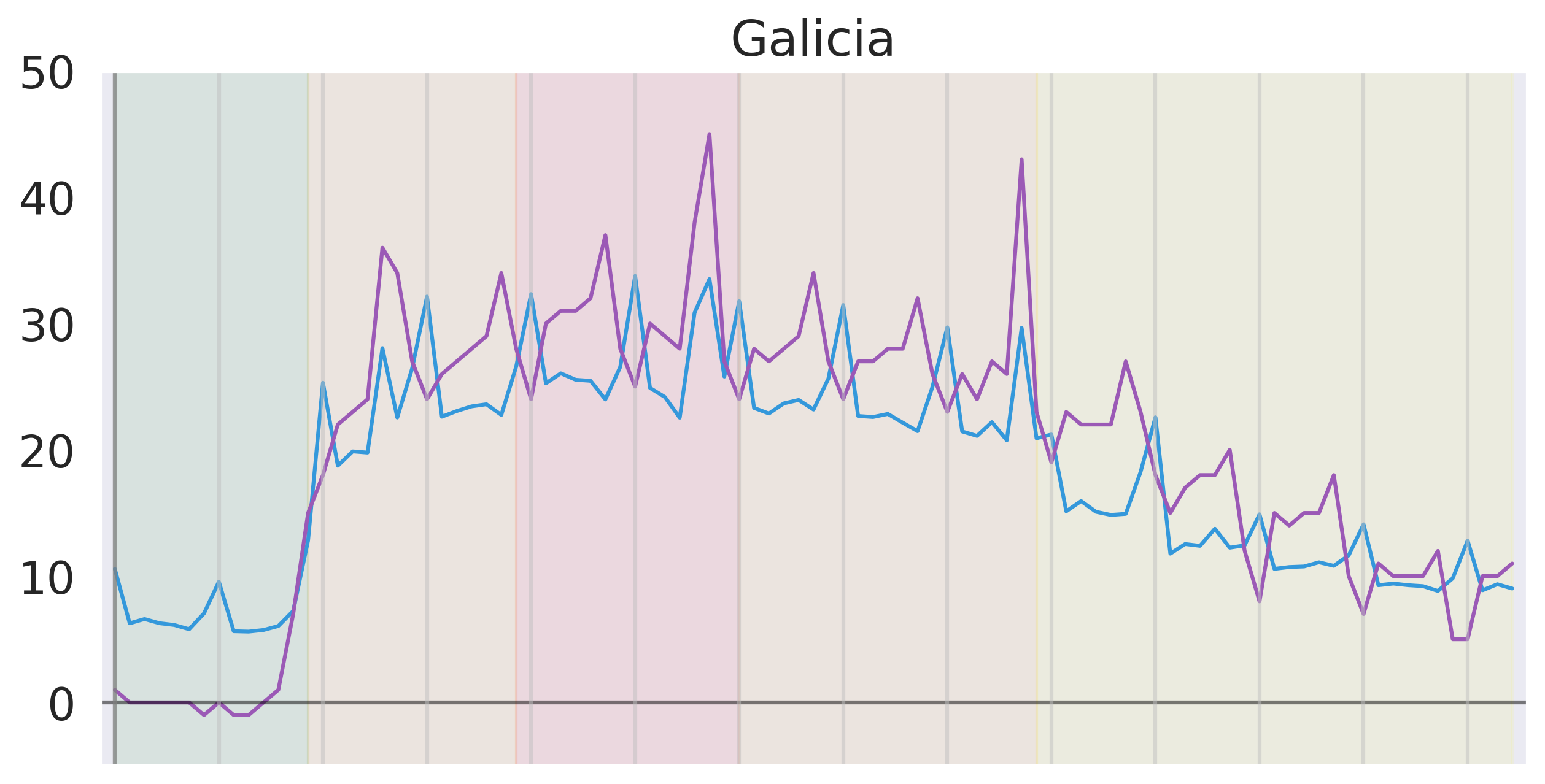}}
    \subfigure{\includegraphics[width=.32\textwidth]{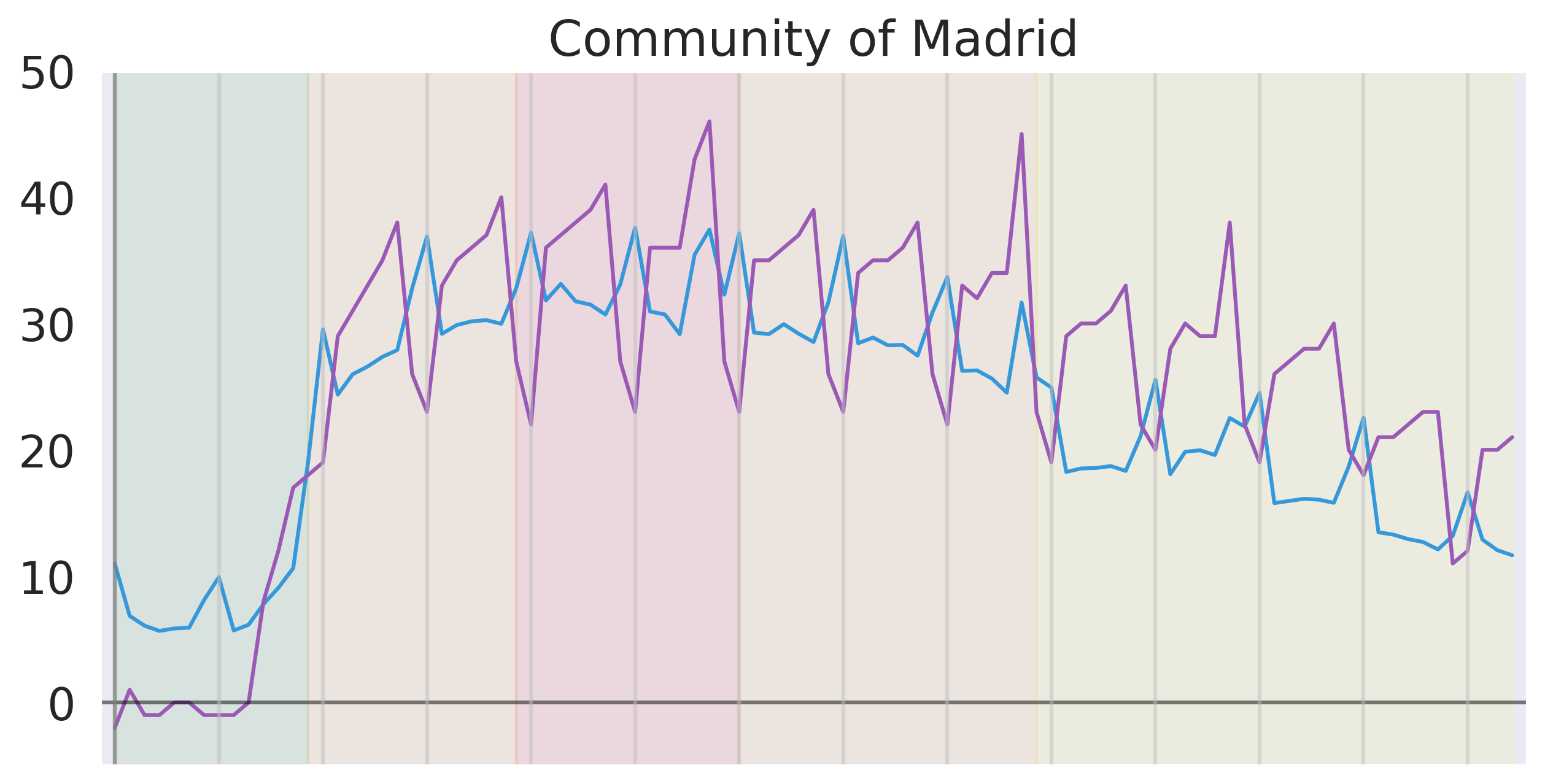}}
    \subfigure{\includegraphics[width=.32\textwidth]{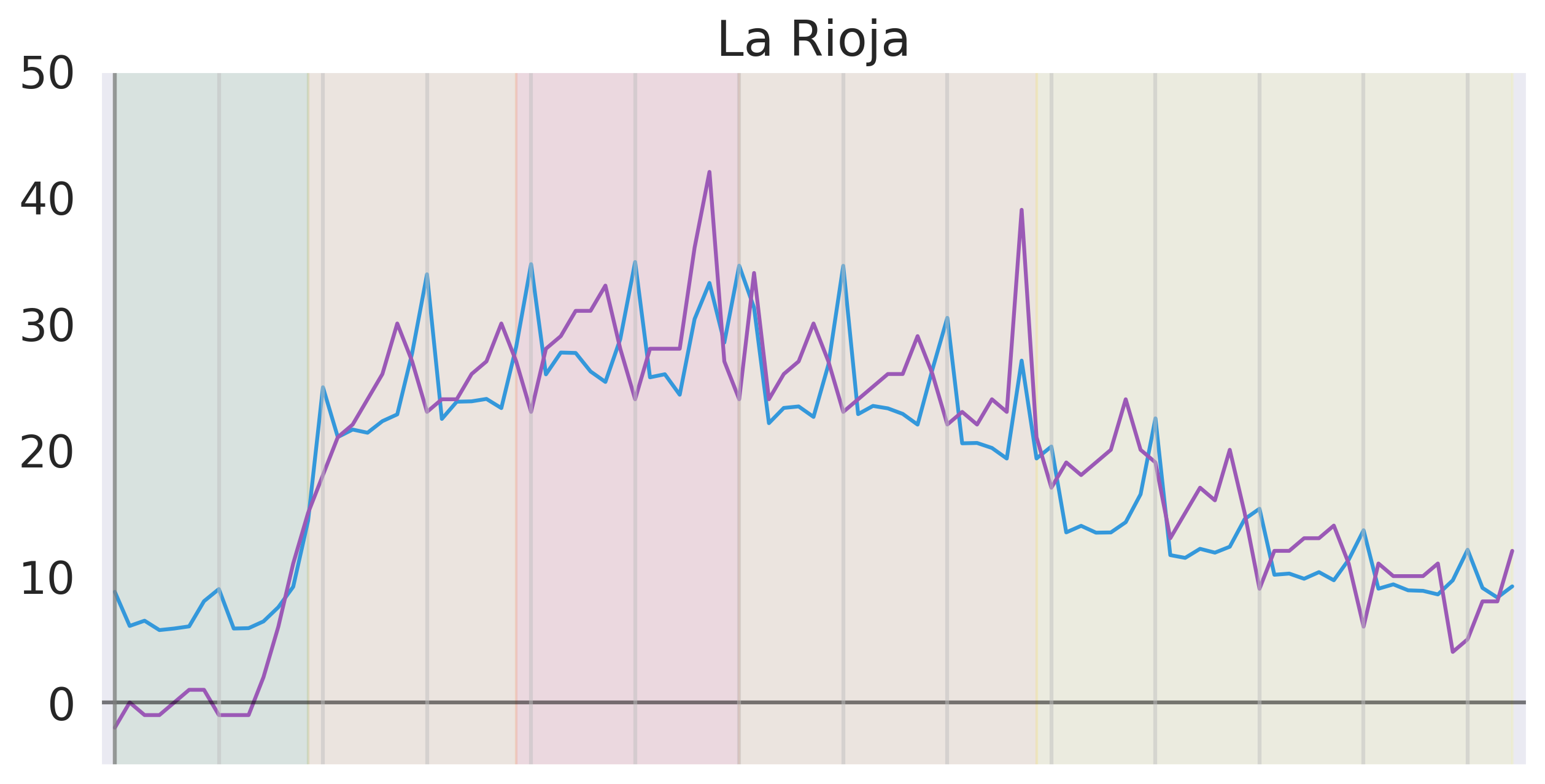}}

    \subfigure{\includegraphics[width=.32\textwidth]{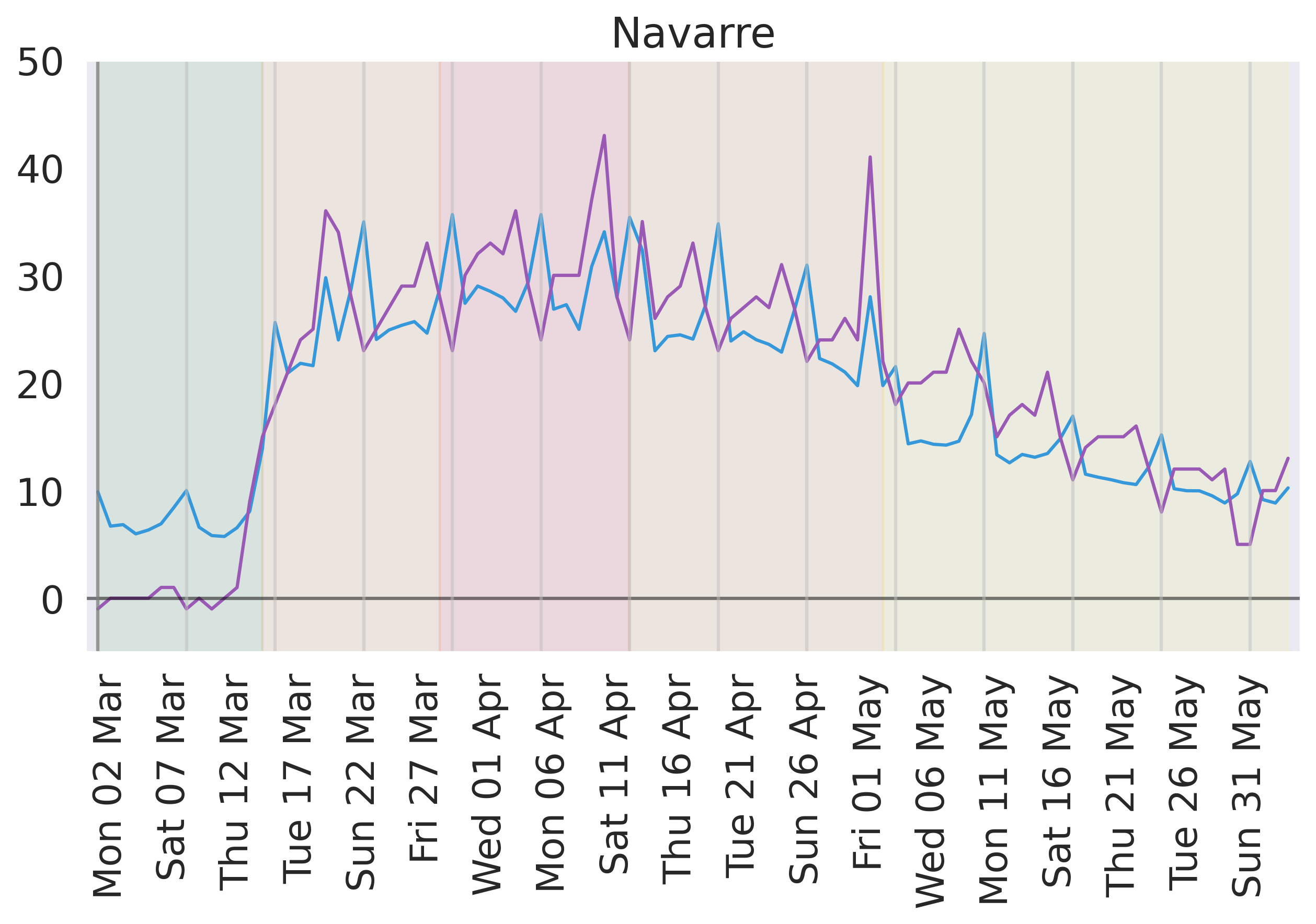}}
    \subfigure{\includegraphics[width=.32\textwidth]{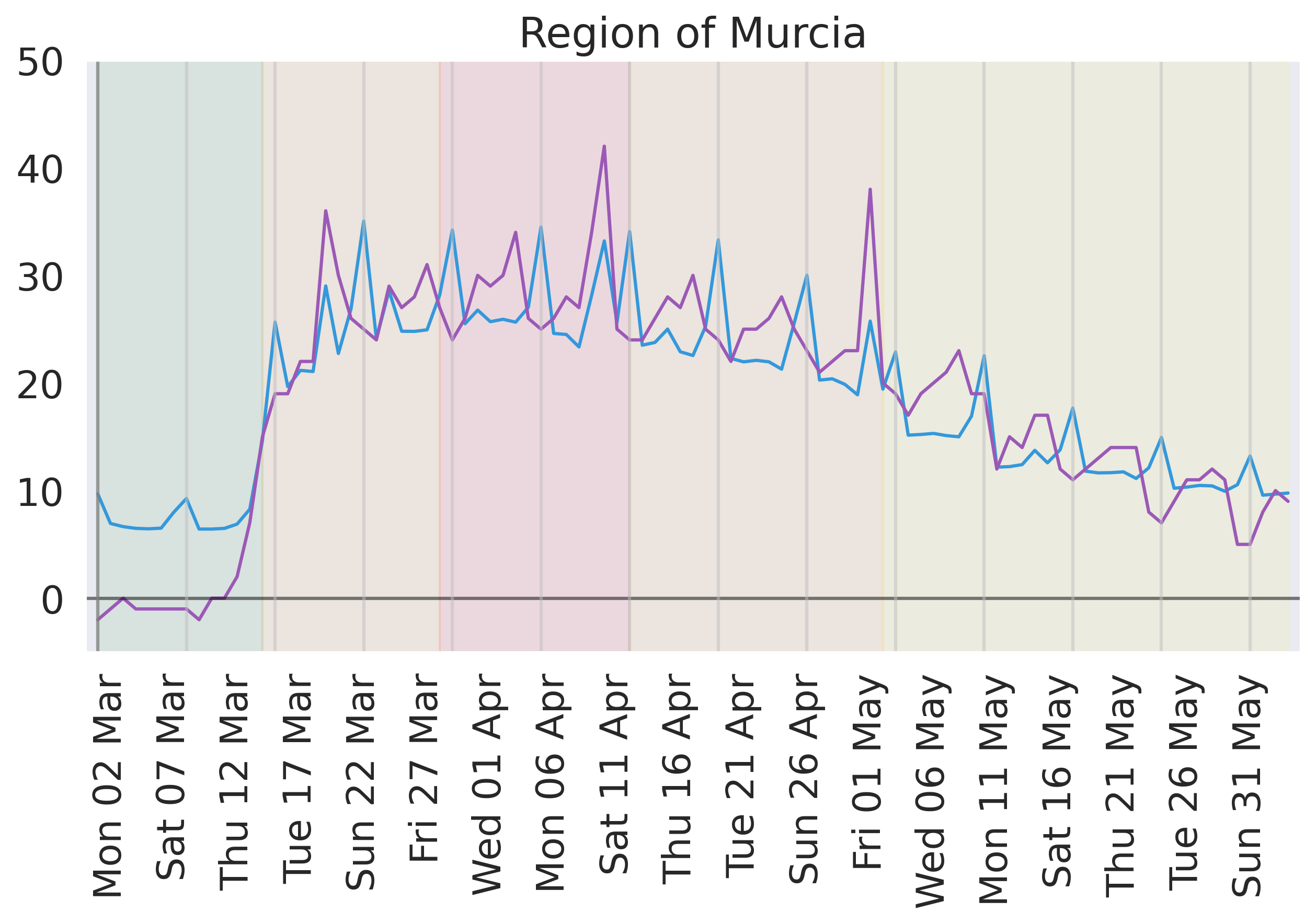}}
    \subfigure{\includegraphics[width=.32\textwidth]{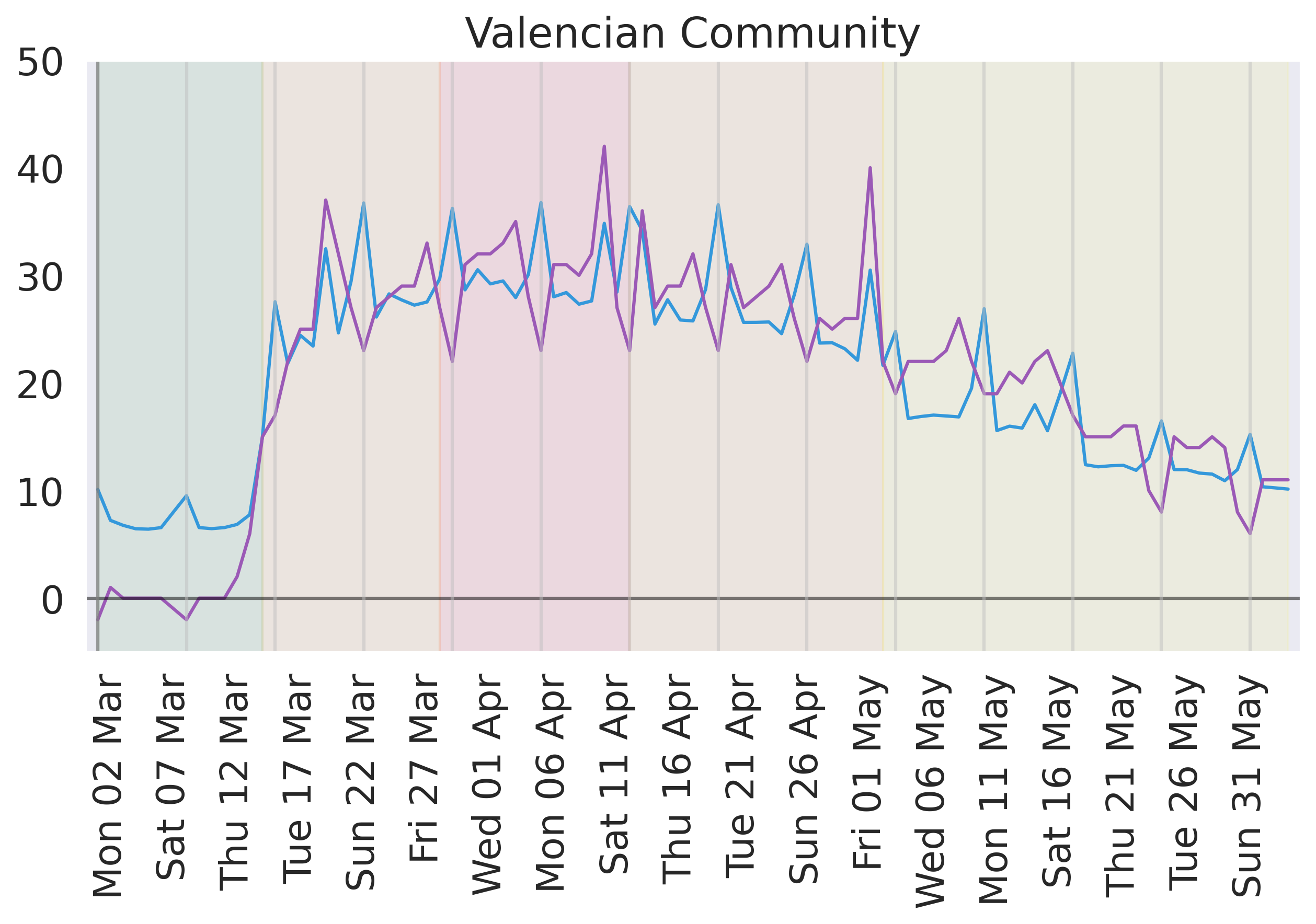}}
        
\caption{Evolution of mobility according to Facebook (blue) and Google (purple) indices, from March to May. The vertical bands (green, orange, red, orange, green) correspond to pre-confinement, state of alarm declaration and lockdown, hard-lockdown, lockdown and de-confinement stages. Grey vertical lines are aligned with every Sunday.}
\label{fig:monthly_trends_1}
\clearpage
\end{figure*}

Our first analysis is focused on the general mobility trend of Spanish regions during the peak of the pandemic. For this, we focus on the three months around the peak of the pandemic in Spain, as shown in Figure~\ref{fig:monthly_trends_1}. These are March, April and May. In this period, mobility in Spain went through at least 5 clear stages (the green, orange, red, orange and yellow bands in Figure~\ref{fig:monthly_trends_1}):

\begin{enumerate}
    \item March 1 - March 13 (2 weeks). A period previous to the declaration of state of alarm, when mobility is expected to be normal. That would mean around zero on the Google index, since this is a measure relative to normality.
    \item March 14 - March 27 (2 weeks). A period of mobility containment after the declaration of the state of alarm and the establishment of a general lockdown.
    \item March 28 - April 12 (2 weeks). A period of maximum mobility containment, what we call \textit{hard lockdown}, after the government reinforced restrictions forbidding all movements not related with essential services.
    \item April 13 - May 1 (3 weeks). General lockdown remains in place, but reinforced policies are lifted (mobility to the workplace is allowed).
    \item May 2 - June 20 (7 weeks). A period of convergence towards new normality, as de-confinement measures are deployed: Kids are allowed outside for limited time, some stores are allowed to open under certain conditions, etc. This process is progressive, with new lifted restrictions every 2 weeks. It goes on until June 21, when the state of alarm ended and new normality began. 
\end{enumerate}

First of all, let us remark the lack of significant differences among regions with regards to the general trend. All show the same overall behavior through time. Mild differences exist in the degree of confinement, and on the recovery speed during de-confinement. \textit{Madrid} for example reaches a higher level of confinement and recovers mobility much slower than \textit{Extremadura}. The main distinctive factor of regions comes from the occurrence of periodic peaks in the data (sometimes upwards, sometimes downwards). We discuss those in detail on the following section.


According to both mobility indicators, the Spanish society assumed and implemented the general lockdown in a matter of 48 hours (from March 12 to March 14). This level of mobility restrain was sustained for seven weeks. For reference, Wuhan, the source of the pandemic, held its lockdown for approximately eight weeks \cite{Reuters-2020-Chinascramblestoc}. Spain mobility was minimized from March 15, with the declaration of the state of alarm, to May 2, with the approval of de-confinement measures (consecutive orange, red and orange bands in Figure~\ref{fig:monthly_trends_1}). After May 2, restrictions were gradually lifted, causing a progressive recovery of mobility that officially ended on June 21.

Within the seven lockdown weeks, two were under hard-lockdown. Since restrictions were enforced by police, mobility in this two week is a good estimate of the maximum mobility restriction that can be held in Spain while keeping essential services running. 

\begin{figure}[!b]
\centering
\includegraphics[width=.9\linewidth]{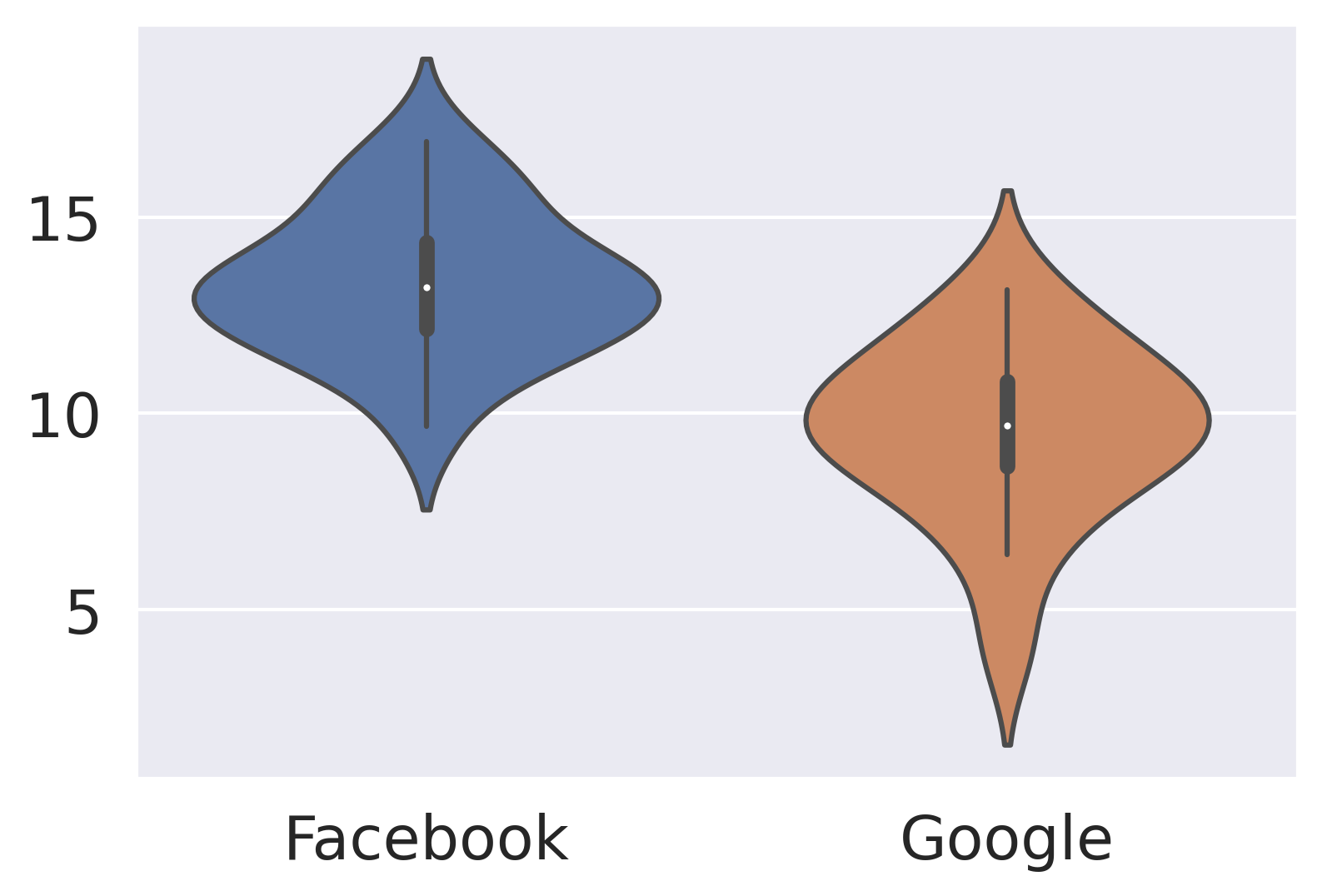}
\caption{Percentual increase in mobility reduction during the hard-lockdown, with respect to mobility during regular lockdown. Distribution includes the values from all 17 Spanish regions.}
\label{fig:violin_hard_lock}
\end{figure}

\begin{figure*}[!t]
\centering
\includegraphics[width=\textwidth]{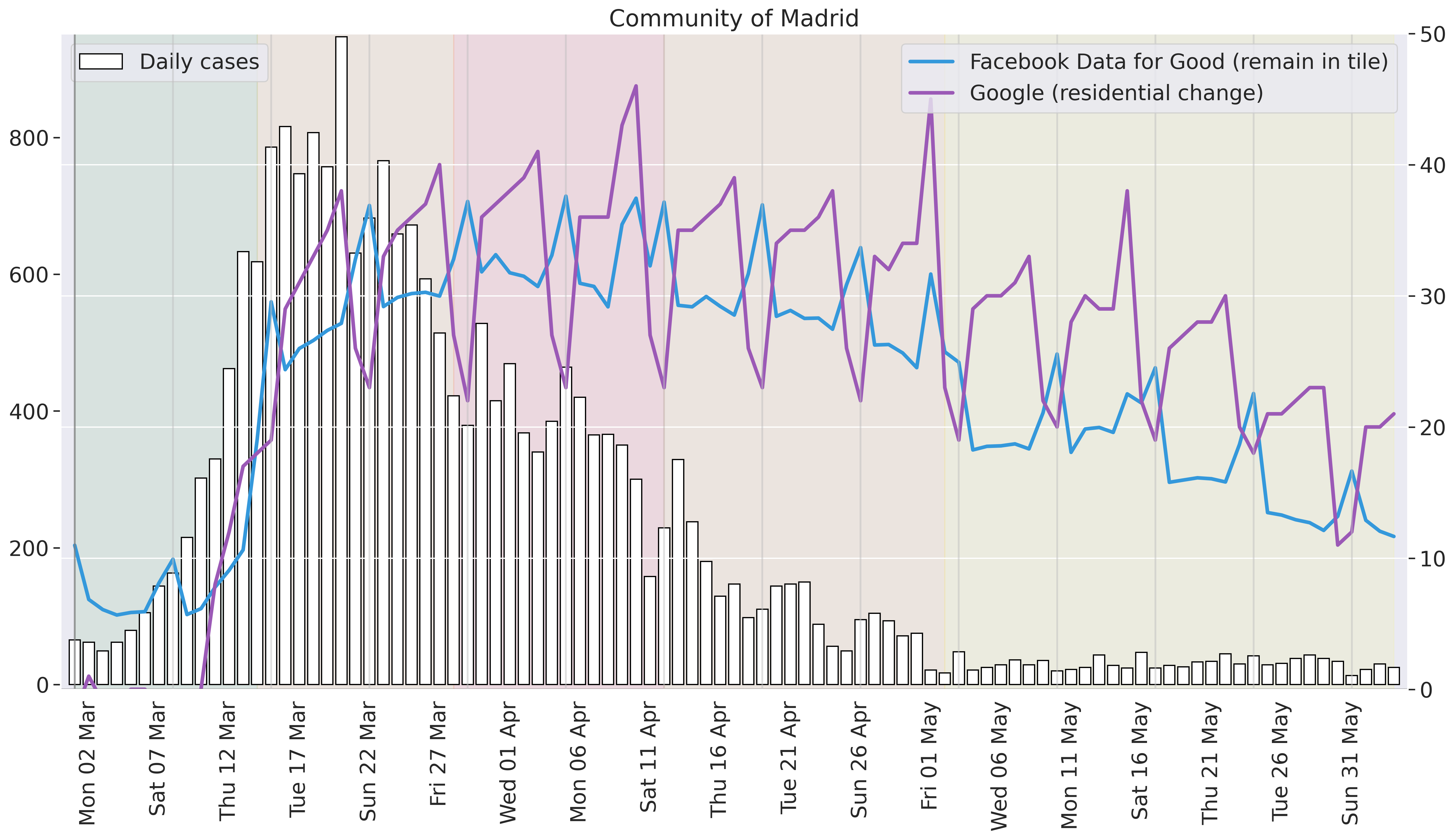}
\caption{Mobility containment in Madrid according to Facebook's remain in tile and Google's residential index, together with the number of reported cases daily in Madrid~\cite{CentroNacionaldeEpidemiologia-unknown-Situacionyevolucio} shifted two weeks early to approximate contagion date.}
\label{fig:trend_with_cases}
\end{figure*}

To further understand the role of the hard-lockdown, we compute an estimate of its impact on mobility. We are interested in its effect when compared to mobility under \textit{regular lockdown}. The regular lockdown includes five weeks of lockdown data (the last two of March and the last three of April) during which traveling to industry and construction workplaces was allowed. This sort of lockdown is assumed to have a less damaging effect on the economy, but it enables infection among co-workers. To compare mobility between both periods we measure the corresponding area under the curve. The higher the area, the higher the constraint. For each period, we normalize the area by the number of weeks. The results for all 17 regions are shown in Figure~\ref{fig:violin_hard_lock} as a separate distribution for the Google and Facebook indices. In this case, Facebook is the most interesting source, since it is an absolute measure and allows us to measure volume of people. This data indicates an increase between 10\% and 15\% in the people who stayed put. The relevance of that number for the containment of the pandemic is unknown to us. \ie we do not know which would have been the evolution of the pandemic if hard-lockdown had not been implemented. One may argue that many more people would have been infected, since regular lockdown enables transmissions on the workplace. However, the number of detected cases did not show any change after hard-lockdown ended, it continued to decrease at a similar rate. One may also argue that the hard-lockdown had a psychological effect on society, boosting resiliency to confinement. Looking at how mobility recovers right after hard-lockdown is lifted, this seems to be a valid hypothesis. If that were the case, a state of alarm without hard-lockdown may have lost adherence faster.

To observe the effect and timing of the policies implemented, we can compare it with the status of the pandemic. For that purpose we use the number of daily reported cases, plotting it against the mobility curves. To approximate the date of the contagion from the date of report, we shift this data two weeks early. This is motivated by current estimates
\cite{Lauer-2020-Theincubationperio}, which assert that the vast majority of COVID-19 patients develop symptoms (if any) before day 14 after contagion. Figure~\ref{fig:trend_with_cases} shows this comparison, for the case of \textit{Madrid}~\cite{CentroNacionaldeEpidemiologia-unknown-Situacionyevolucio}, the region with the most cases and the strongest lockdown adherence. 

In Figure~\ref{fig:trend_with_cases} we observe the beginning of the lockdown overlapping with the initial containment of the pandemic. That is, the number of contagions halts its exponential trend around the date of the state of alarm declaration. Although we do not know the exact role of the state of alarm, it is shown to be a clearly correlated factor. The seven weeks of general lockdown coincide with the seven weeks of strongest pandemic rate reduction. That is the time it took the pandemic to reach a basal situation in the region, with less than 100 reported cases a day. According to this estimate, this situation may have been reached around the starting date of the de-confinement process. If all this was the case, the duration of the Spanish lockdown (7 weeks) was a very good fit to the evolution of the pandemic. A shorter lockdown may have induced a significantly higher risk of relapse, and a longer one seems unnecessary in sight of the contagion numbers estimated to be taking place in early May. Let us remark that during the crisis policy makers could only use current daily cases for their decision making. That is, without the 2 week shift we performed in Figure~\ref{fig:trend_with_cases}. The de-confinement process was therefore a bolder (and riskier) initiative than Figure~\ref{fig:trend_with_cases} illustrates.

\subsection{Peaks}\label{sec:peaks}

The general trend described in the previous section is explained by the different stages of confinement enforced by the Spanish government. On top of this trend, we can see the occurrence of a number of peak values, happening periodically and on all regions. These peaks occur mostly on Sundays (marked with grey vertical lines in Figure~\ref{fig:monthly_trends_1}), and to a lesser degree on Fridays and Saturdays. Let us discuss Sundays and Fridays in detail, since Saturdays seem to be a middle ground, transitioning between both.

Sunday peaks are anticorrelated in terms of relative mobility (Google goes down) and absolute mobility (Facebook goes up). This means the number of people moving is small when compared to near-by days of confinement. It also means that the number of people moving is not so different from what it used to be, when compared to the same day of the week on normality. In other words, people were not moving much on Sundays before the COVID-19 crisis, and during it they were moving less. Accordingly, even though the decrease on mobility on Sundays is not as big as on other days of the week, it still accounts for the day of the week with less absolute mobility. That would make Sundays the best candidates for the mobility of risk population.

In contrast, Friday peaks exhibit a rather different behavior. In this case, the relative mobility decreases sharply (Google goes up), while the absolute mobility remains stable or decreases mildly (Facebook goes flat or slightly up). This indicates Fridays are the days with the most different mobility patterns with respect to the previous normality (relative change), which speaks of the high mobility taking place on a normal Friday. On these days is when society is showing its biggest change, leveling mobility to the rest of the working week. That would make Fridays the best candidates for communication and support (\eg mental health assessment). Let us remind the reader that these insights may be reinforced by the bias in the data, which favours the presence of the young segments of society.

\begin{figure}[!b]
\centering
\includegraphics[width=.99\linewidth]{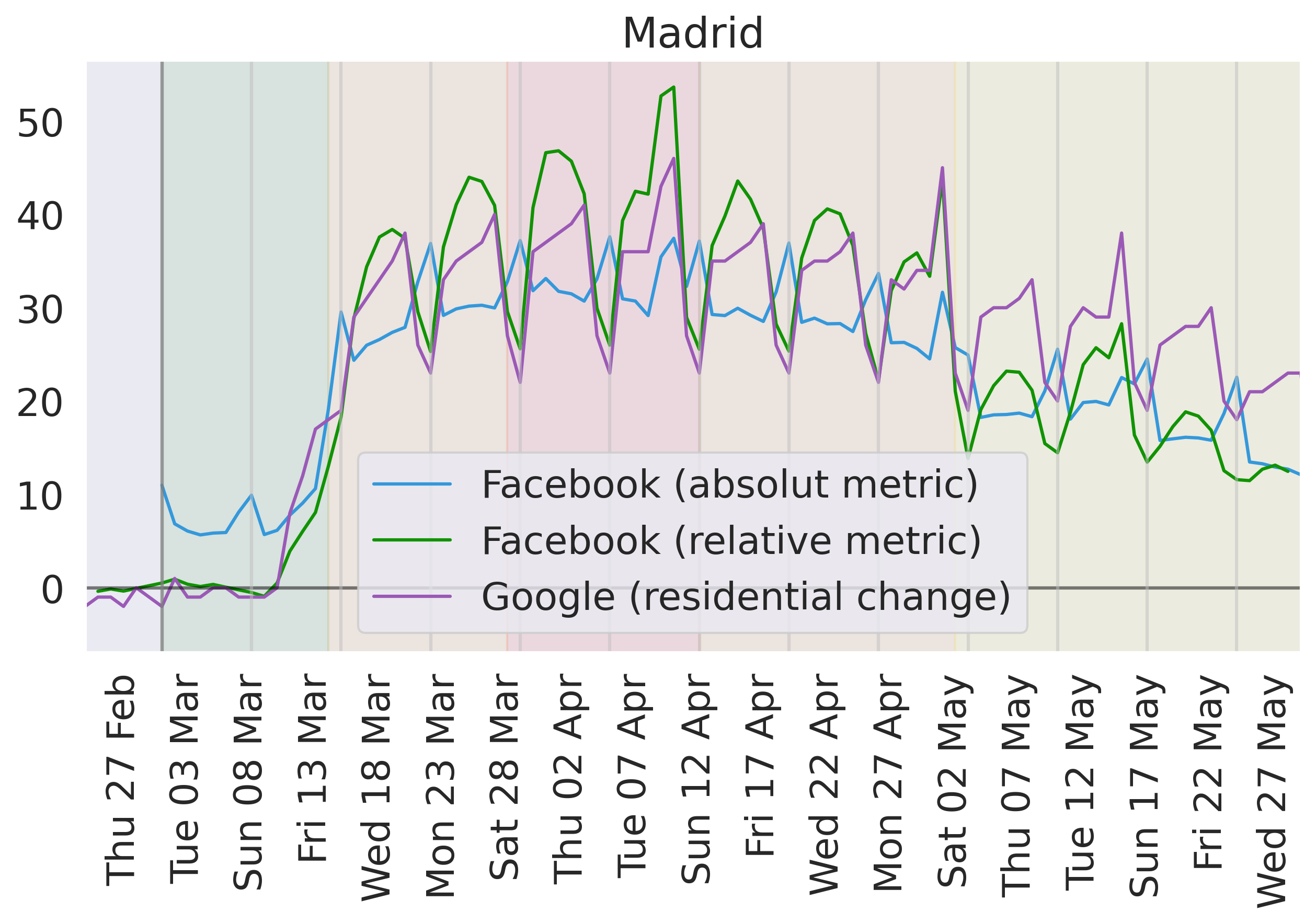}
\caption{Evolution of mobility according to Facebook absolute change (blue), Facebook relative change (green) and Google (purple). Facebook relative change has been divided by 100 to match the scale of the other two indices. The vertical bands (green, orange, red, orange, green) correspond to pre-confinement, state of alarm declaration and lockdown, hard-lockdown, lockdown and de-confinement stages.  Grey vertical lines are aligned with every Sunday. }
\label{fig:facebook_relative}
\end{figure}

Let us now conduct an experiment to validate the hypothesis that peaks are related to the relative or absolute nature of measures. We transform the Facebook measure from an absolute one to a relative one, using as a baseline analogous remain in tile data, from February 24 to March 8, before the first regional restriction measures. This baseline is computed weekday-wise, like Google's. The result is shown in Figure~\ref{fig:facebook_relative}, together with the original Facebook data, and the Google measure. The first obvious change are Sundays, which now peak downwards like Google. In fact, our relative Facebook measure perfectly aligns with Google around weekends (Friday to Monday) during the whole lockdown. This may be caused by differences in the data (both data sources have different resolutions to measure movement), or by differences in the baselines given the daily consistency.

\begin{figure*}[th]
\centering
\includegraphics[scale=0.3]{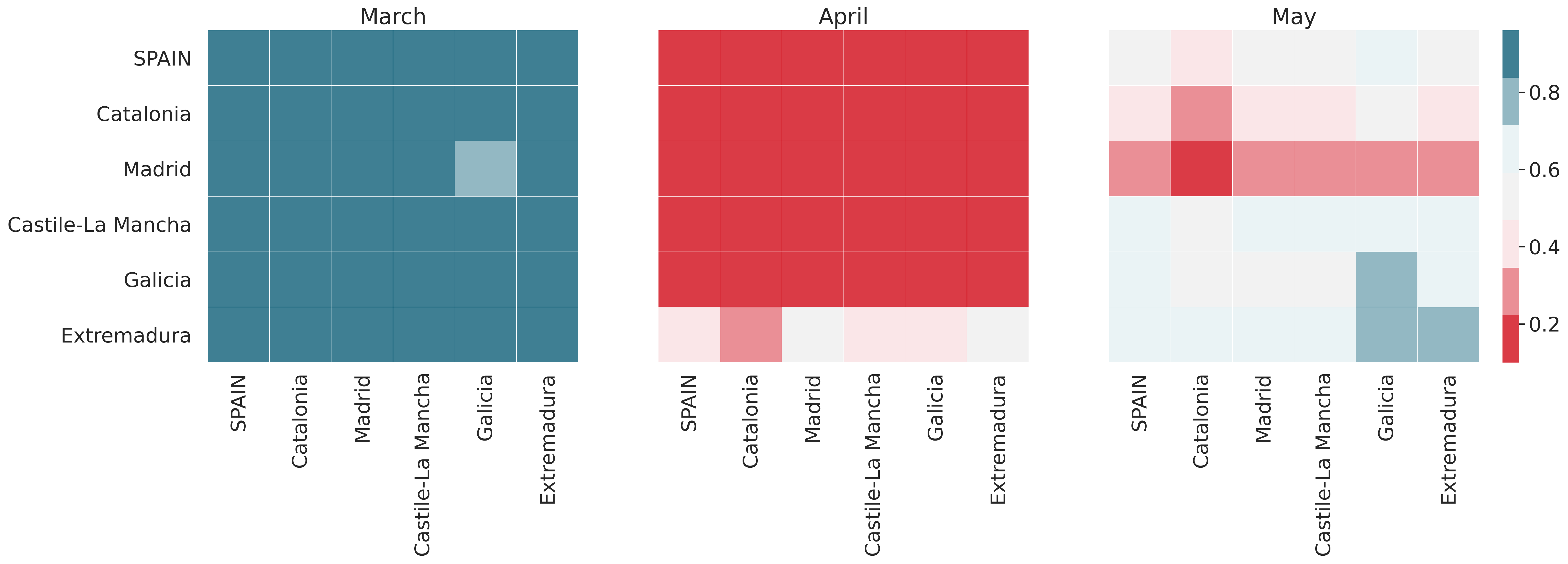}
\caption{Pearson correlation between Facebook's remain in tile and Google's residential index for different regions and months. Facebook plotted on the horizontal axis, and Google on the vertical one.}
\label{fig:correlation}
\end{figure*}

Understanding the nature of these peaks is important because of the effect these may have on certain metrics. As shown in Figure~\ref{fig:correlation}, Pearson correlation between Facebook's remain in tile and Google's residential index varies significantly from month to month. On March, mobility exhibited a very clear trend as a result of the establishment of confinement measures. In this setting, the correlation between both indices is clear (around 0.9 Pearson correlation on average), and the peaks are not disruptive enough as to alter it. On the other hand, mobility during April was stable, as the whole month was under lockdown. This entails an overall flat behavior of the indicators. A context in which the inverted peaks have a dramatic effect, destroying all correlation between indices. Finally, May appears as a middle ground. There is a generalized mobility trend, which reduces the upsetting effect of the peaks, but the trend is not strong enough as to completely overpower the noise.\dgnote{This paragraph is as good as it gets without really being science.}

\begin{figure*}[th]
\centering
\includegraphics[scale=0.46]{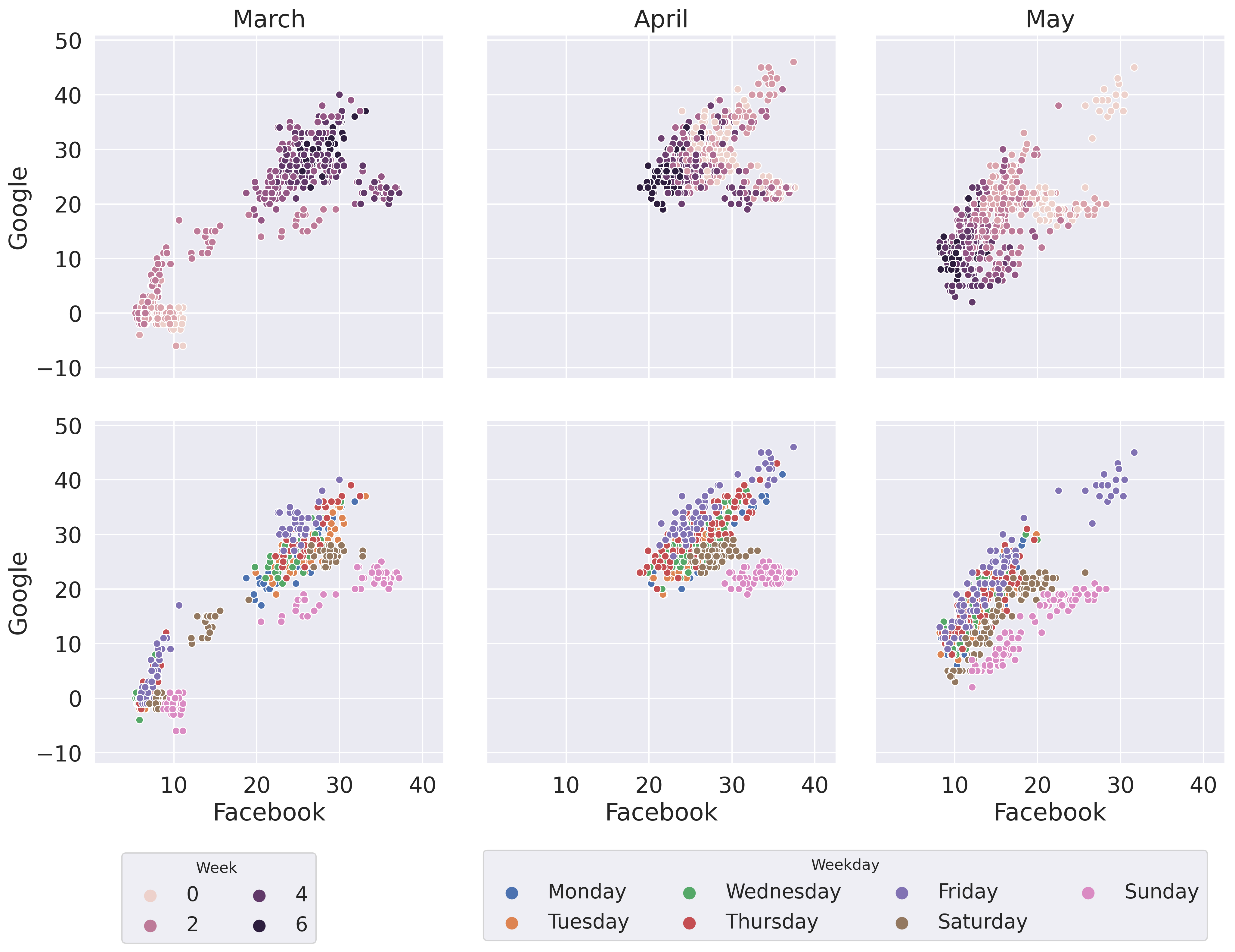}
\caption{Evolution of mobility for Facebook (horizontal axis) and Google (vertical axis) data. Each dot represents a single day in a single region. First column holds data for March, second for April and third for May. First row shows the data with a color gradient indicating the week of the month. Second row shows the same data colored based on day of the week.}
\label{fig:day_dots_month}
\end{figure*}

\subsection{Daily Trend}\label{sec:daily_trend}

Week days have an important role in the characterization of mobility. Let us now study the same data, but this time from the perspective of days. To do so we plot the Facebook and Google mobility indices as two different axis. Figure~\ref{fig:day_dots_month} provides two visualizations for the first three months of the pandemic in Spain. On the top row, the color gradient shows the change through time, week by week. On the bottom row, week days are color coded to illustrate the differences between days. In these plots, horizontal axis show absolute change (the more to the right, the bigger number of people stay at home) while the vertical axis shows relative change (the more up, the more percentage of people stay at home with respect with normal instances of that day).

The first visible thing in Figure~\ref{fig:day_dots_month} is the correlation between both values, as all data is mostly gathered around the diagonal. The top row shows the evolution of mobility, starting from the axis origin (bottom left) and suddenly jumping to the top right quarter of the plot as lockdown is implemented. The last Friday and Saturday before the lockdown (second week of March) are the only days in the middle of that jump. During confinement (April) data is rather stable in that area, until the de-confinement measures (May) bring it down and left again, but this time in a slow manner.

The visualization using both Google and Facebook as axis shows the clear correlation between them. In general, as relative mobility increases/decreases, so does absolute mobility. However, this relation seems to be somewhat dependant on the day of the week. As shown on the bottom row of Figure~\ref{fig:day_dots_month}, Sundays have a rather different behavior in terms of relative mobility (it shows less affection in this metric) while Friday represents the opposite (it shows more affection in relative mobility). This is a different visualization of the same phenomenon observed in the peaks of Figure~\ref{fig:monthly_trends_1} and discussed in \S\ref{sec:peaks}.

\subsection{The New Normality}\label{sec:new_norm}

On the second half of May, Spain started to lift the confinement measures that had been in place in the country for two months \cite{Hegedus-2020-ECMLCoviddashboard}. The process was asymmetrical, with regions with better pandemic indicators (\ie number of daily cases, number of available hospital beds, \etc) de-confining faster than others. Detailed maps of the differential treatment of regions can be found in governmental sources \cite{PresidenciadelGobiernodeEspana-2020-PlanparalaTransic} This process ended on June 21, when the state of alarm (and all mobility restrictions) was lifted. On that date, the whole country officially entered the new normality.
\dgnote{ToDo (not for arxiv version) new normality vs old normality plot. 4 colors 1 box?}

\begin{figure*}[th]
\centering
\includegraphics[width=\textwidth]{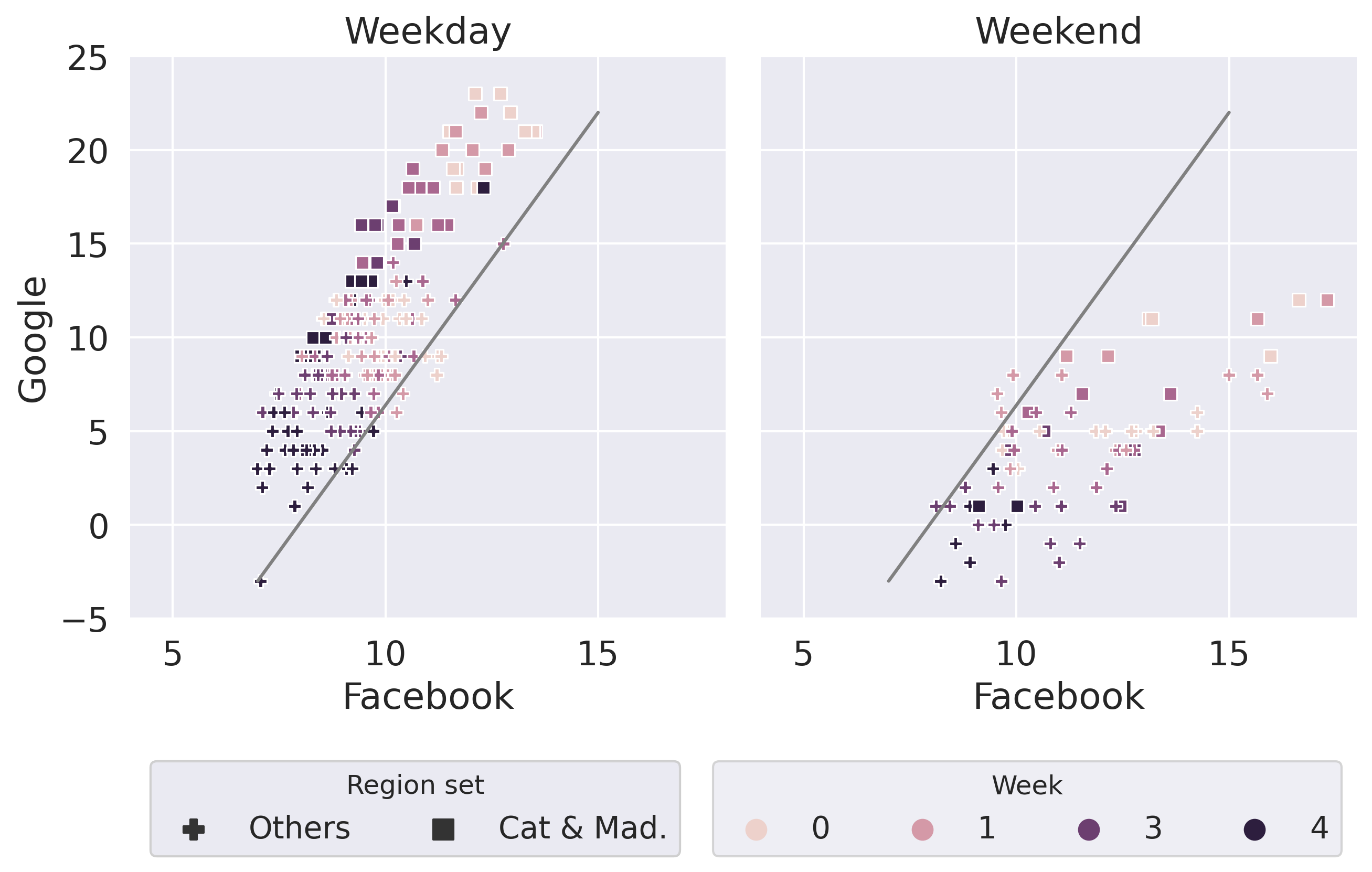}
\caption{Mobility values from Monday to Friday (left plot) and from Saturday to Sunday (right plot). Color gradient indicates the week number (from March 25 to June 27), and the shape of the marker indicates the region set (Crosses: Asturias, Navarre, La Rioja, Region of Murcia, Extremadura and Galicia. Squares: Catalonia and Madrid). Grey diagonal line is the same on both plots, used for the purpose of visual reference.}
\label{fig:new_normality_short}
\end{figure*}

\begin{figure}[h!]
\centering
\includegraphics[width=.4\textwidth]{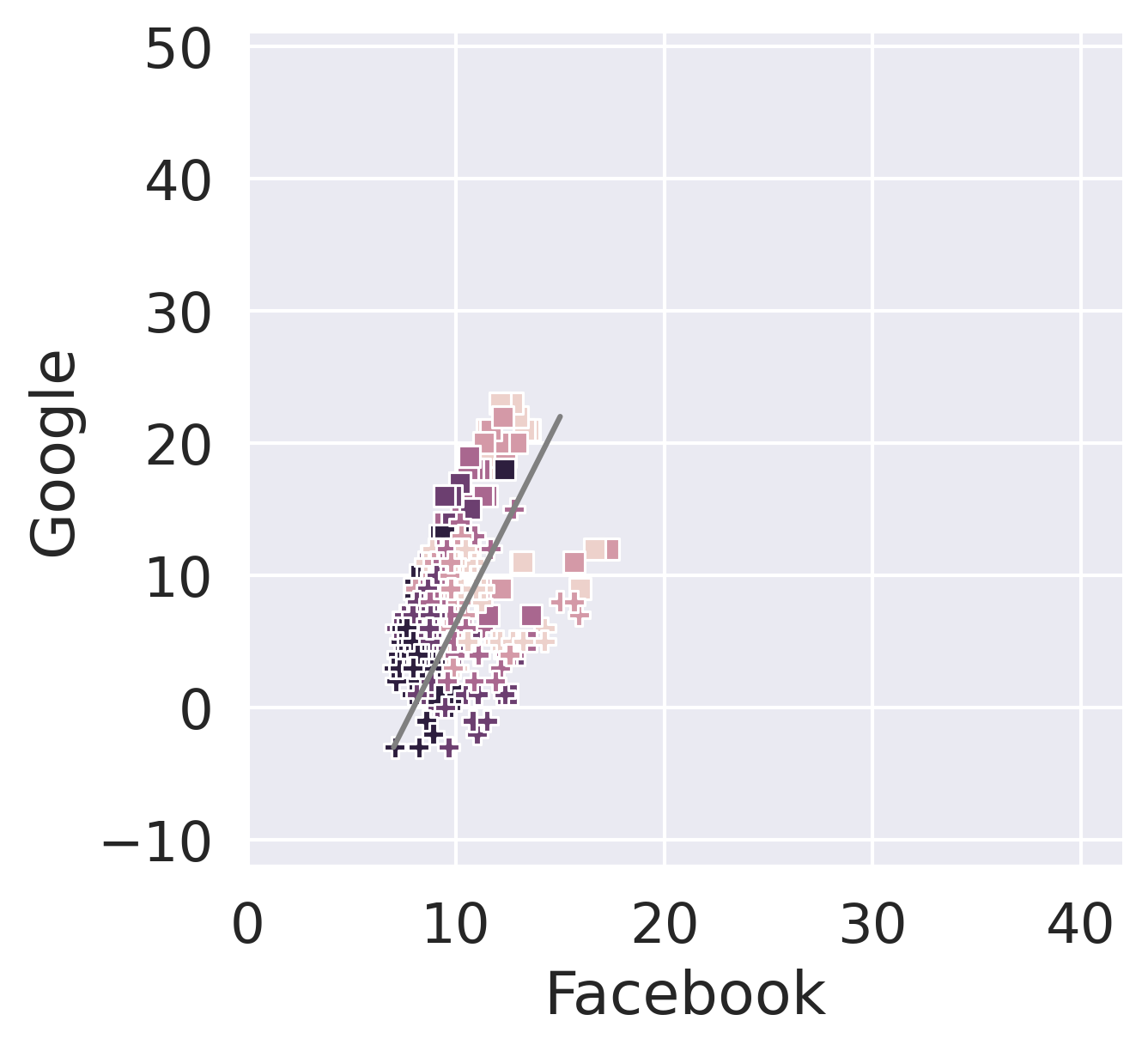}
\caption{Mobility from March 25 to June 27, using the same scale of Figure~\ref{fig:day_dots_month}. The data of this figure is zoomed in and split between working days and weekend days in Figure~\ref{fig:new_normality_short}.}
\label{fig:new_normality_original}
\end{figure}

Figure~\ref{fig:new_normality_short} includes the last four weeks of state of alarm (but without a generalized lockdown), and the first one of new normality. To facilitate visualization we change zoom in and the axis scale with respect to Figure~\ref{fig:day_dots_month}. Nonetheless, to enable comparison with the rest of the period under study, in Figure~\ref{fig:new_normality_original} we plot the same data using the scale used in Figure~\ref{fig:day_dots_month}

The progression of mobility towards the axes origin is still visible and this smaller scale as weeks go by (in color gradient), for both working days (Monday to Friday) and weekend days (Saturday and Sunday). To compare the new normality with the old one, we must focus on the Google axis, since this is relative to a baseline (January 3 to February 6). On the weekends mobility is already at Google baseline levels, with all values between -5 and 5 on the last week (the new normal one). In contrast, working days are showing a higher difference with respect to the baseline, with several values in the last week between 5 and 15 in the Google axis. This indicates that the change implemented by the Spanish society during the new normality is focused on working days, while weekends are back to how they were.

The recovery of old normality is not homogeneous amount regions either. Catalonia and Madrid, the regions with the biggest metropolitan areas, clearly lag behind. Asturias, Navarre, La Rioja, Region of Murcia, Extremadura and Galicia are way ahead. Of those, only Murcia has a population density above 100 (132 people/km${^2}$), hinting a potential relation with this indicator. The lag of Catalonia and Madrid during the first 4 weeks is likely related with the fact that these regions were slightly behind in the removal of restriction measures. However, during the 5th and last week of data all regions were under the same conditions, and Catalonia and Madrid still exhibit higher levels of mobility reduction. This may be related with the role of large metropolitan areas, were it is harder to keep a safe distance, and with the fact that both regions reported the highest absolute volume of infections during the pandemic. Both these factors are strong psychological enablers of self-responsibility, which may have an effect of adherence to mobility reduction during the new normality.

\section{Conclusions}\label{sec:concl}

In this work we consider the use of private data sources (Google and Facebook) for assessing the levels of mobility in a country like Spain. By doing so, we draw conclusions on two fronts. First, on the behavior and particularities of private data sources. And second, on how mobility changed during the COVID-19 pandemic in Spain.

Regarding private data sources, we have shown the differences between using an absolute measure (like Facebook) and a relative measure (like Google). Both of them have limitations when used in isolation. The former lacks a contextualization of its values, while the latter depends entirely on the baseline used. When used together, they provide a visualizing of mobility where consistent patterns can be easily identified (as presented later in this section). For specific purposes, using a single data source may suffice, as long as it fits the goal:

\begin{itemize}
    \item An absolute measure like Facebook's can be very useful for epidemiologic purposes, as it provides an pure measurement of mobility. That includes estimating number of contacts in a society, modeling the spread of the virus, and measuring the impact of policies on absolute mobility.
    \item A relative measure like Google's can be very useful for socio-economic purposes, as it provides a contextualized measurement of mobility. That includes understanding the change caused by the new normality, and the economic impact of mobility restriction policies.
\end{itemize} 

On the second topic of this paper, the analysis of Spanish mobility during the COVID-19 pandemic, we extract several conclusions. On one hand, data shows a huge mobility containment, sustained for a month and a half (March 15 to May 1st, approximately), very close to its theoretical limit (as represented by mobility during the hard-lockdown). This duration was sufficient to contain the spread of the virus and bring infection numbers down to traceable scale. In hindsight, the policies implemented in Spain seem appropriate and proportional to the severity of the situation. That being said, the role, timing and convenience of the hard-lockdown remains to be further discussed. Our work shows a relatively modest contribution of this policy to mobility reduction. On the other hand, the hard-lockdown may have had an effect on prolonging adherence.

Our work identifies mild differences between regions during the three months of restricted movement. Certain regions had a stronger adherence to confinement than others, mostly in relative terms. This may be caused by regional differences in pre-pandemic mobility, which is used as baseline for the relative measurement. A similar artifact are the inverted peaks of weekends, where a relative measure spikes down and an absolute measure spikes up. As demonstrated, this the result of combining a measure relative to the weekday with an absolute measure. 

We also saw significant differences among days. Weekends exhibit the highest volume of mobility reduction in absolute terms, even during the hard-lockdown, when traveling to work was forbidden for all except essential services. At the same time, weekends have the smallest mobility reduction in relative terms, indicating that the effort society had to make in this regard with respect to its previous patterns was smaller. Fridays and Sundays are particularly relevant days, the first because it represents the biggest change from normal behavior, the second because it represents the biggest absolute decrease in mobility. These particularities could be exploited for the general good.

Finally, we analyzed the new normality by looking at the weeks of de-confinement, up until June 27, a week after the state of alarm was lifted on the whole of Spain. In this period, we found Saturdays and Sundays to be already at pre-pandemic levels of mobility. In contrast, working days (Monday to Friday) still show significant differences. The new normality also shows differences between regions, particularly for working days. Regions with large metropolitan areas exhibit a reduction in mobility between 4\% and 14\% after restrictions were lifted. Indeed, the new normality is most new on urban working days.
    

\section*{Acknowledgements}
We would like to thank Facebook and Google for releasing the data that made this work possible. We also appreciate the insights and support of Amaç Herdağdelen, Alex Pompe and Alex Dow on the interpretation of peaks. Part of this work was done under the Global Data Science Project for COVID-19. We would also like to thank Daniel López-Codina, Sergio Alonso, and Clara Prats for fruitful discussions. Finally, part of this research has received funding from the European Union’s Horizon 2020 Programme under the SoBigData++ Project, grant agreement num. 871042.

\bibliography{main.bib}
\bibliographystyle{ieeetr}

\end{document}